\documentclass{aa}
\usepackage{graphicx}
\usepackage{txfonts}
\usepackage{rotating}
\usepackage{natbib}
\usepackage{colortbl}
\usepackage{longtable}
\bibpunct{(}{)}{;}{a}{}{,} 

\newcommand{\micron}{\hbox{$\mu{\rm m}$}}                      

\newcommand{\msun}{\mbox{$M_{\odot}$}}
\newcommand{\Mdot}{\mbox{$\dot M$}}
\newcommand{\Menv}{\mbox{$M_\mathrm{env}$}}
\newcommand{\Pdot}{\mbox{$\dot P$}}
\newcommand{\Tbol}{\mbox{$T_\mathrm{bol}$}}
\newcommand{\Lbol}{\mbox{$L_\mathrm{bol}$}}
\newcommand{\Lsun}{\mbox{$L_{\odot}$}}
\newcommand{\Fco}{\mbox{$F_\mathrm{CO}$}}

\newcommand{\Bin}{\mbox{\hbox{Bin}}\,}

\defcitealias{paperI}{Paper~I}
\defcitealias{class}{Paper~II}
\defcitealias{outflows}{Paper~III}

\begin{document}
   \title{Star formation in Perseus}

   \subtitle{III. Outflows}

   \author{J. Hatchell\inst{1}, G. A. Fuller\inst{2},
   J. S. Richer\inst{3}}
   \authorrunning{Hatchell et al.}
   \titlerunning{SCUBA Perseus survey -- III. Outflows}
   
   \offprints{hatchell@astro.ex.ac.uk}

   \institute{School of Physics, University of Exeter, Stocker Road, Exeter EX4 4QL, U.K.
              \and School of Physics and Astronomy, University of Manchester, P.O. Box 88, Manchester M60 1QD, U.K.
              \and Cavendish Laboratory, Cambridge CB3 0HE, U.K.
              }

   \date{}
   
   \abstract{We present a search for outflows towards 51 submillimetre
     cores in the Perseus molecular cloud.  }
   {Our first objective is to identify the protostellar population
     through the detection of molecular outflows.  Our second aim is
     to consistently derive outflow properties from a large
     homogeneous dataset within one molecular cloud in order to
     investigate further the mass dependence and time evolution of
     protostellar mass loss.}
   {We used the James Clerk Maxwell Telescope to map $2'\times 2'$
     regions around each core in $^{12}$CO~3--2.  Where molecular
     outflows were detected we derived momentum fluxes.}
   {Of the 51 cores, 37 show broad linewings indicative of molecular
     outflows.  In 13 cases, the linewings could be due to confusion
     with neighbouring flows but 9 of those sources also have
     near-infrared detections confirming their protostellar nature.
     The total fraction of protostars in our sample is 65\%.  All but
     four outflow detections are confirmed as protostellar by Spitzer
     IR detections and only one Spitzer source has no outflow, showing
     that outflow maps at this sensitivity are equally good at
     identifying protostars as Spitzer.  Outflow momentum flux correlates both
     with source luminosity and with core mass but there is
     considerable scatter even within this one cloud despite the
     homogeneous dataset.  We fail to confirm the result of Bontemps
     et al.~(1996) that Class~I sources show lower momentum fluxes on
     average than Class~0 sources, with a KS test showing a
     significant probability that the momentum fluxes for both
     Class~0s and Class~Is are drawn from the same distribution.}
   {We find that outflow power may not show a simple decline between
     the Class~0 to Class~I stages.  Our sample includes low momentum
     flux, low-luminosity Class~0 sources, possibly at a very early
     evolutionary stage.  If the only mass loss from the core were
       due to outflows, cores would last for
       $10^5$--$10^{8}$~years, longer than current estimates of
       $1.5$--$4\times 10^5$ years for the mean lifetime for the
       embedded phase.  Additional mechanisms for removing mass from
     protostellar cores may be necessary.  }


%


\maketitle

\section{Introduction}
\label{sect:introduction}

Submillimetre (submm) continuum surveys of the Perseus molecular
cloud, such as our SCUBA survey, (\citealt{paperI}, hereafter Paper~I,
see also \citealt{kirk06}), and the similar 1300\micron\ survey with
Bolocam \citep{enoch06}, have identified over 100 submm cores.  These
cores could be either protostellar or starless in nature: protostellar
cores can be identified by radio or infrared detection of a central
source, or evidence for a protostellar outflow.  In a sister paper
\citep[ hereafter Paper~II]{class} we classify the protostellar cores
on the basis of their spectral energy distributions, including the new
data from Spitzer \citep{c2d,jorgensen07}.  Here we take a
complementary approach and look for the presence of molecular outflows
in order to separate the protostellar cores from the starless
population.  As outflow activity is thought to begin almost as soon as
the formation of a hydrostatic core, outflows are a good indicator of
the presence of embedded protostars.


In addition, studying molecular outflows is an important way to get
information on the accretion/mass ejection process which drives
protostellar evolution.  Several important studies have shown that the
outflow force or momentum flux \Fco\ is correlated with both mass and
luminosity of the protostellar core \citep{cabritbertout92,batc96}, a
correlation which recently has been shown to hold over several orders
of magnitude \citep{churchwell97,beuther02,zhang05} and held to be strong evidence
for a common protostellar formation mechanism for high- and low-mass
sources alike.  The details of mass accretion and ejection, however,
remain uncertain with competing mechanisms for the interaction between
mass flow and magnetic fields in the inner parts of accretion disks
\citep{shu94,wardle93,pelletier92} and considerable uncertainty
on how the accretion rate
varies with time.  One certainty, however, seems to be that the
accretion rate and the momentum output in the jet/wind are strongly linked
\citep{shu94,ferreira95,batc96}, and therefore, assuming a roughly
constant mass entrainment fraction, one can use the outflow
momentum as a
surrogate measure of the momentum in the jet/wind, and therefore the mass
accretion rate onto the protostar.

We therefore set out to make small outflow maps around the submm
cores in Perseus to determine their outflow properties.  With a
consistent dataset for a moderately large sample drawn from one
molecular cloud, we can look
statistically at the outflow detection rate and the variation of
the momentum flux with other parameters.  

We assume a distance of 320~pc for the Perseus molecular cloud based on the
Hipparcos distance of IC348 \citep{dezeeuw99}, consistent with our
earlier paper \citepalias{paperI}, though readers should be aware that
other recent studies \citep{kirk06,enoch06} have assumed a closer
distance of 250~pc based on extinction studies
\citep{cernisstraizys03} and references therein).

In Sect.~\ref{sect:obs} we describe the observations.
Sect.~\ref{sect:results} contains the results and an examination of the detection statistics.  Momentum fluxes and outflow evolution are considered in Sect.~\ref{sect:pflux} .  In Sect.~\ref{sect:lifetimes} we discuss the derived outflow
mass loss rates and the implications for core destruction timescales
given the core masses implied by the submm emission
\citepalias{paperI,class}.   Finally, our conclusions
are given in Sect.~\ref{sect:summary}.


\section{$^{12}$\hbox{CO} 3\hbox{--}2 outflow observations}
\label{sect:obs}

Outflow maps were made in $^{12}\hbox{CO }J=3\hbox{--}2$ in the years
2000--2003 using RxB3 on the James Clerk Maxwell Telescope (JCMT).
$^{12}$CO~3--2 is a better tracer of molecular outflows than lower-J
transitions as it is more sensitive to warm outflow than cold ambient
emission.  The sample consisted of contiguous areas around the known
clusters B1, IC348, L1448, L1455, NGC1333 \citep{kneesandell00}, and
a number of isolated submm cores.  The map areas were extended
$2'\times 2'$ around each isolated source or tiled to cover the
cluster regions.  The submm cores which lie within the outflow maps
are listed in Table~\ref{tbl:pflux}.  Positions and further
information on these cores can be found in \citetalias{paperI} and
\citetalias{class}.  The beamsize at 345~GHz is $14''$ and the maps were
fully sampled at $5''$ spacing in RA and Dec.  A correction for main
beam efficiency of $\eta_{\mathrm MB} = 0.63$ was applied.  The noise
level was 0.5~K or less on 625~kHz ($0.54 \hbox{ km s}^{-1}$)
channels.  The weather conditions were various and resulting system
temperatures were 350--500~K.  


The $^{12}$CO~3--2 data for NGC1333 were published in
\citet{kneesandell00}, and kindly made available by the authors.  The
spatial sampling of the NGC~1333 observations was $10''$ rather than
the $5''$ of the more recent observations.  In order to
match the other datasets as closely as possible for analysis, the maps
were resampled onto a $5''$ grid.  This was done by interpolating with
a $14''$ Gaussian truncated at $40''$.  This results in a map with the
same intensity ($T_{MB}$) scale but with $20''$ resolution.

The sample (Table~\ref{tbl:pflux}) was selected before the analysis of the corresponding SCUBA
data was complete, and therefore is biased towards the obvious bright
clusters and previously known protostars.  In Fig.~\ref{fig:sample} we
show the outflow sample (marked by stars) on a plot of luminosity vs.
core mass for the entire population of submm cores in Perseus
\citepalias{class}.  The sample is indeed biased towards high
luminosities with no outflow maps of sources below 0.25~\Lsun.  The
proportion of the total number of cores decreases with decreasing
mass.  However, the fraction of $\Lbol > 0.25\Lsun$ cores studied
remains fairly constant except for the very lowest mass sources below
our completeness limit at 10~K of $\Menv < 0.5~\msun$ ($5\sigma$).
For the very highest mass sources, $\Menv > 12\msun$, all sources were
observed as there were no low-luminosity sources in the SCUBA sample.
Thus the most significant bias of the sample is its lack of low
luminosity sources.

Small outflow maps provide a good estimate of the momentum flux
\citep{fullerladd02,mangum98} but, except for very small flows, do not provide
estimates of the full extent, mass, kinetic energy, or dynamical
timescale of outflows.  Our map radius of $1'$ corresponds to 0.1~pc
assuming a distance of 320~pc for the Perseus molecular clouds, or a
dynamical time of 2000~years for an outflow advancing at 50~km/s.

\begin{figure*}[p!]
\centering
\includegraphics[scale=0.9,angle=0]{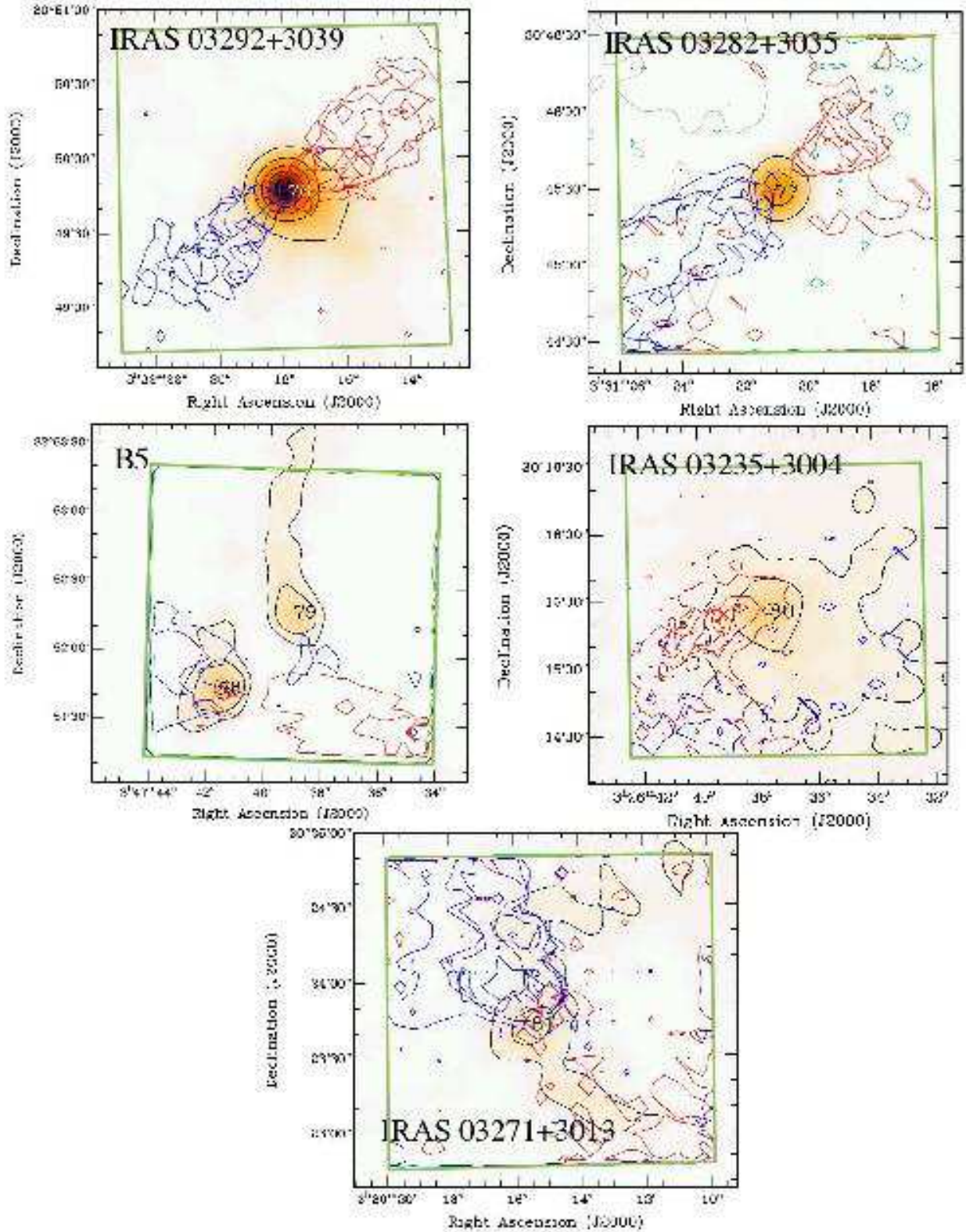}\\
\caption{ $^{12}$CO~3--2 outflow maps of IRAS~03292+3039 (76), IRAS~03282+3035 (77), B5~IRS1 (78) and neighbouring submm source (79), IRAS03235+3004 (80), and IRAS~03271+3013 (81).  Outflows are displayed as a renzogram, equivalent to superimposing a series of channel maps each with one contour at a fixed intensity level \citep[eg. ][]{renzogram}.  Velocity ranges were as follows: (76) 2.0~K~km~s$^{-1}$, -5--4 (blue) and 9--15~km~s$^{-1}$ (red) ; (77) 3.0~K~km~s$^{-1}$,  -5--5 (blue) and  7--15~km~s$^{-1}$ (red); (78) and (79) 2.5~K~km~s$^{-1}$, -5--8 (blue) and 12--15~km~s$^{-1}$ (red) ; (80) 2.0~K~km~s$^{-1}$, -5--3 (blue) and 6--15~km~s$^{-1}$ (red); and (81) 1.0~K~km~s$^{-1}$, -5--6 (blue) and 8--15~km~s$^{-1}$ (red).  Colourscale: 850\micron\ SCUBA map with black contours at 100, 200, 400, 800, 1600~mJy/beam. The area observed in $^{12}$CO~3--2 is outlined in green.}
\label{fig:outflows}
\end{figure*}

\addtocounter{figure}{-1}
\begin{figure*}[p!]
\centering
\includegraphics[scale=0.8,angle=0.0]{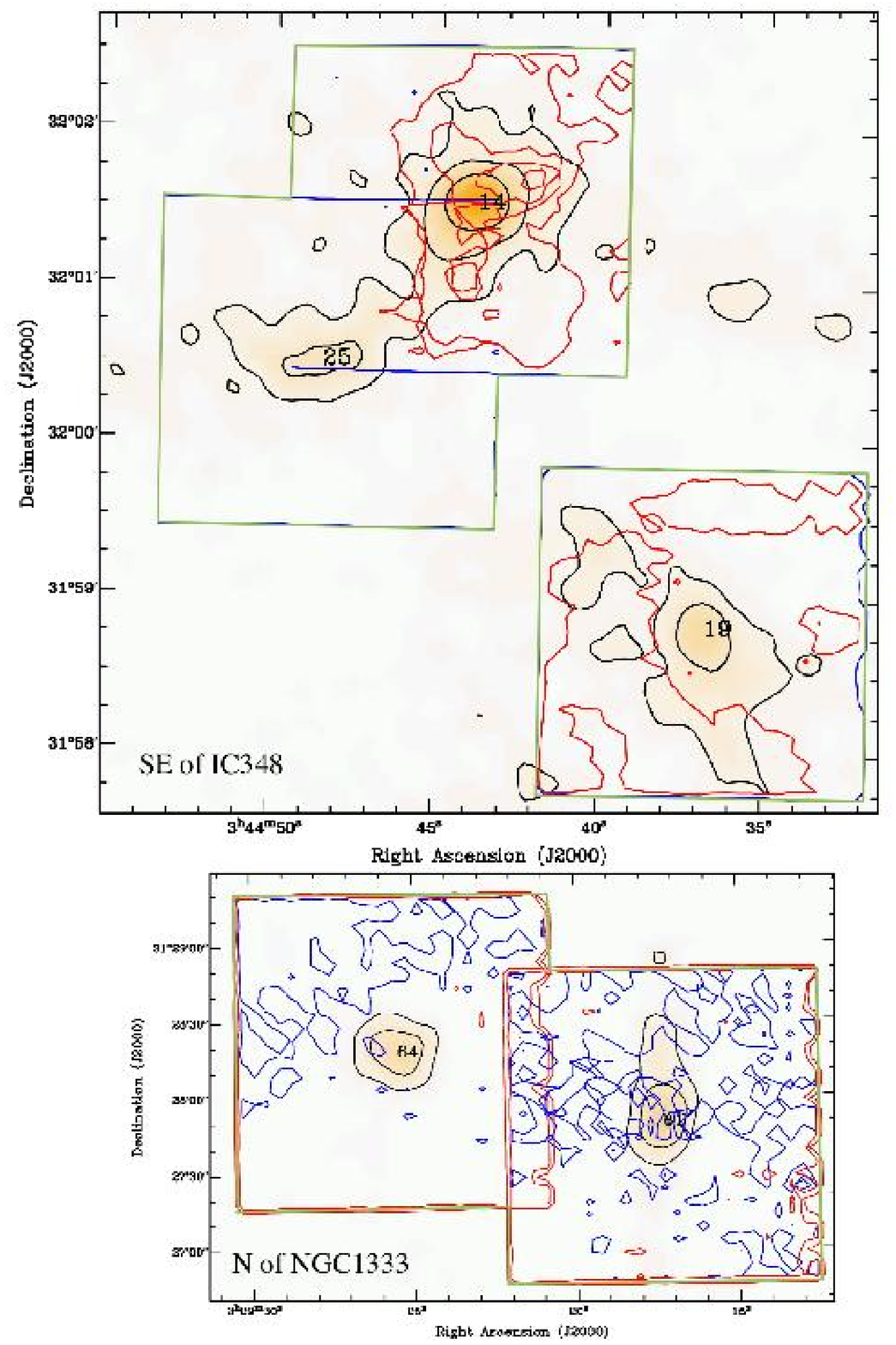}\\
\caption{continued. {\bf Region to the SE of IC348}  The renzogram intensity level is 3~K~km~s$^{-1}$ and included channels from -5--7~km~s$^{-1}$ (blue) and 11--15~km~s$^{-1}$ (red).  {\bf Region in the N of NGC1333.}  The renzogram intensity level is 2~K~km~s$^{-1}$ and included channels from -5--1~km~s$^{-1}$ (blue) and 12--15~km~s$^{-1}$ (red).  850\micron\ contours and colourscale are as before.}
\end{figure*}

\addtocounter{figure}{-1}
\begin{figure*}[p!]
\centering
\includegraphics[scale=0.8,angle=0.0]{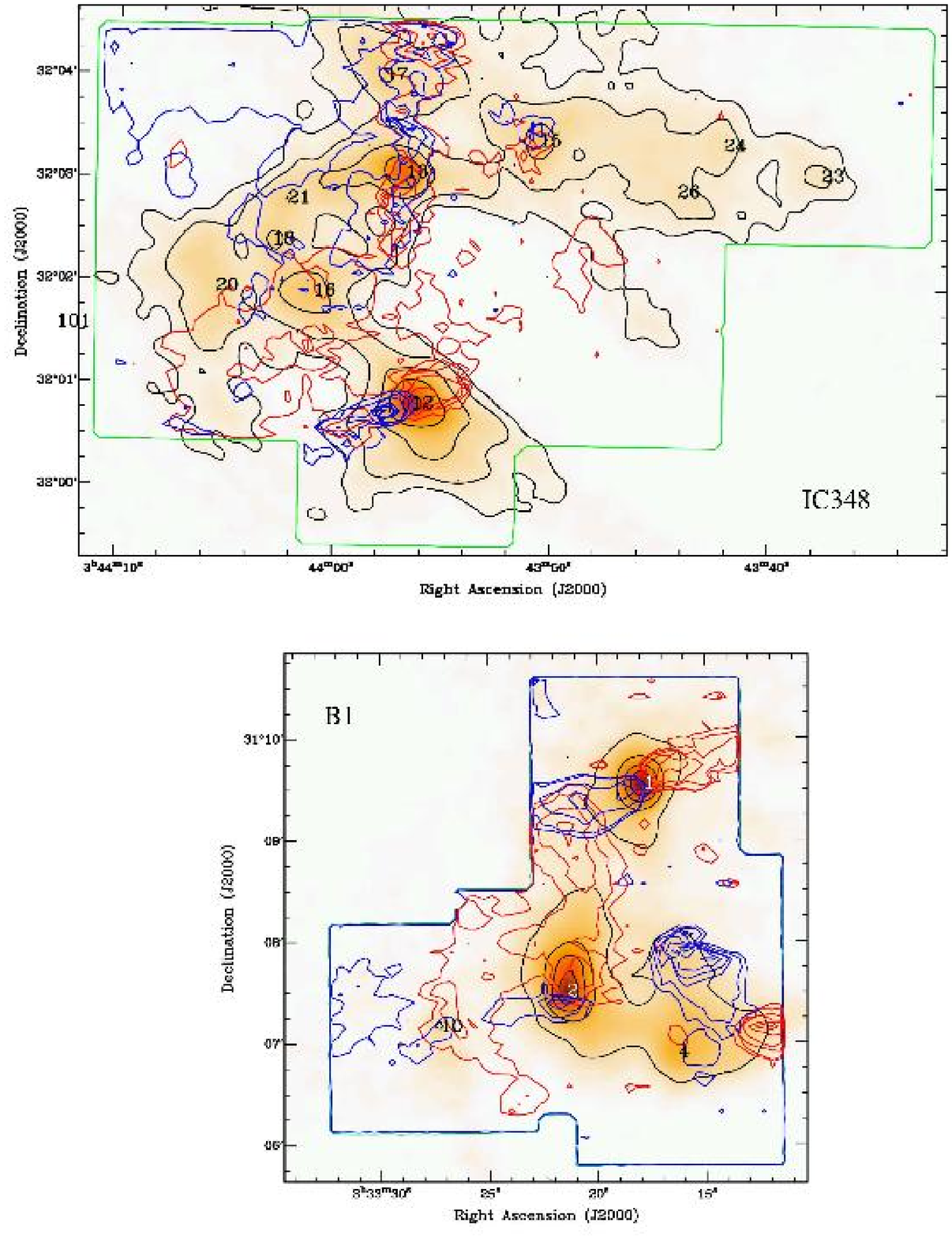}\\
\caption{continued.  {\bf IC348}, region to the S of IC348, including HH211 (12) and IC348~mms (13). The renzogram intensity level is 2~K~km~s$^{-1}$ and included channels from -5--6~km~s$^{-1}$ (blue) and 11--15~km~s$^{-1}$ (red).  {\bf B1} $^{12}$CO~3--2 outflow maps of B1, including B1-c (1), B1-bN/S (2), b1-d (4) and IRAS 03301+3057 (7).  The renzogram intensity level is 3~K~km~s$^{-1}$ and included channels from -5--4~km~s$^{-1}$ (blue) and 9--15~km~s$^{-1}$ (red).  850\micron\ contours and colourscale are as before.}
\end{figure*}

\addtocounter{figure}{-1}
\begin{figure*}[p!]
\centering
\includegraphics[scale=0.80,angle=0.0]{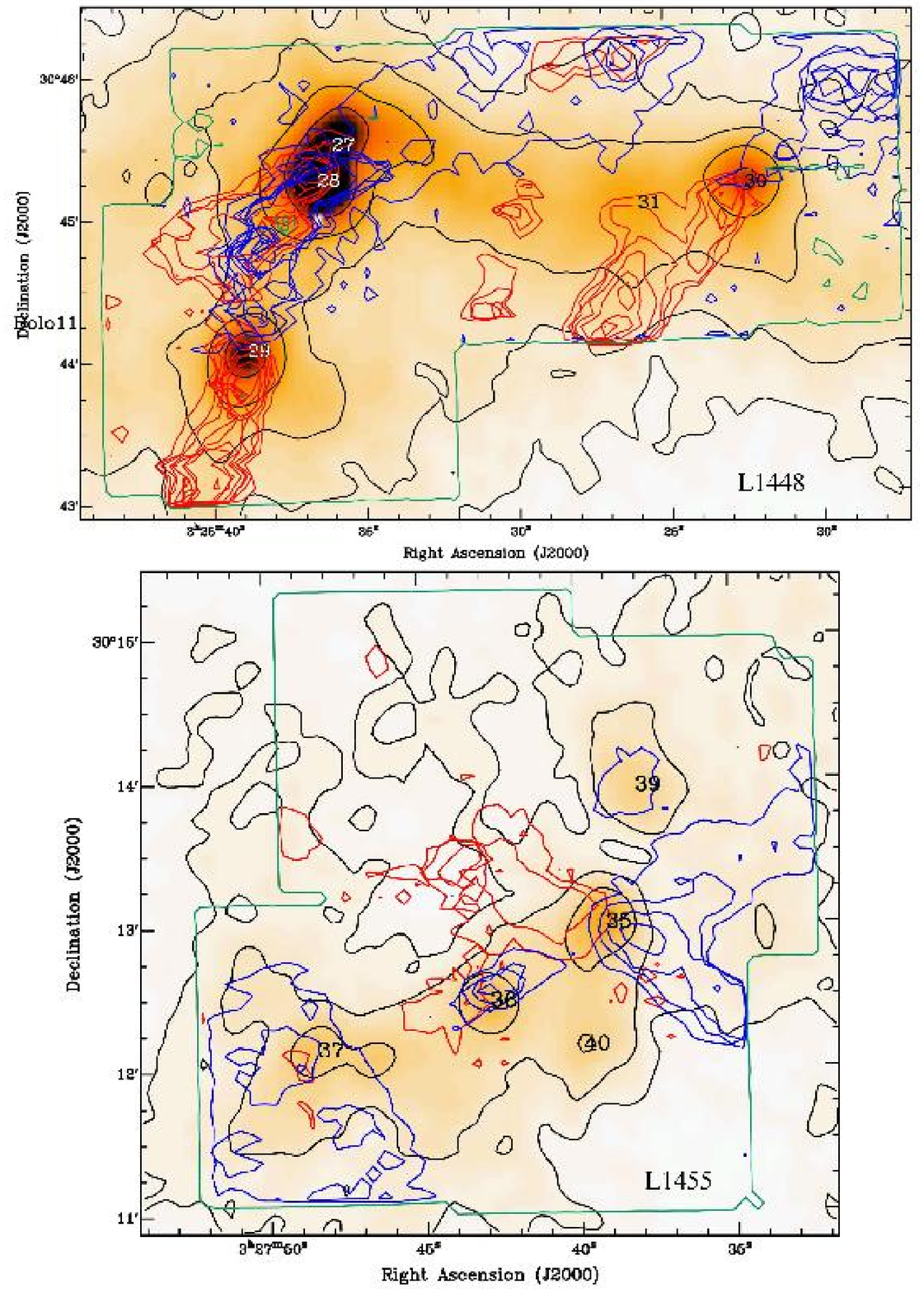}\\
\caption{{\bf continued. L1448} including L1448~N/NW (27), L1448~C (28) and L1448~IRS2 (30).  The renzogram intensity level is 3.0~K~km~s$^{-1}$ and included channels from -5--0~km~s$^{-1}$ (blue) and 8--15~km~s$^{-1}$ (red).  {\bf L1455} including RNO15 (35), PP9 (37), and L1455~FIR2 (39).  The renzogram intensity level is 1.8~K~km~s$^{-1}$ and included channels from -5--2~km~s$^{-1}$ (blue) and 11--18~km~s$^{-1}$ (red).  850\micron\ contours and colourscale are as before.}
\end{figure*}

\addtocounter{figure}{-1}
\begin{figure*}[p!]
\centering
\includegraphics[scale=0.80,angle=0.0]{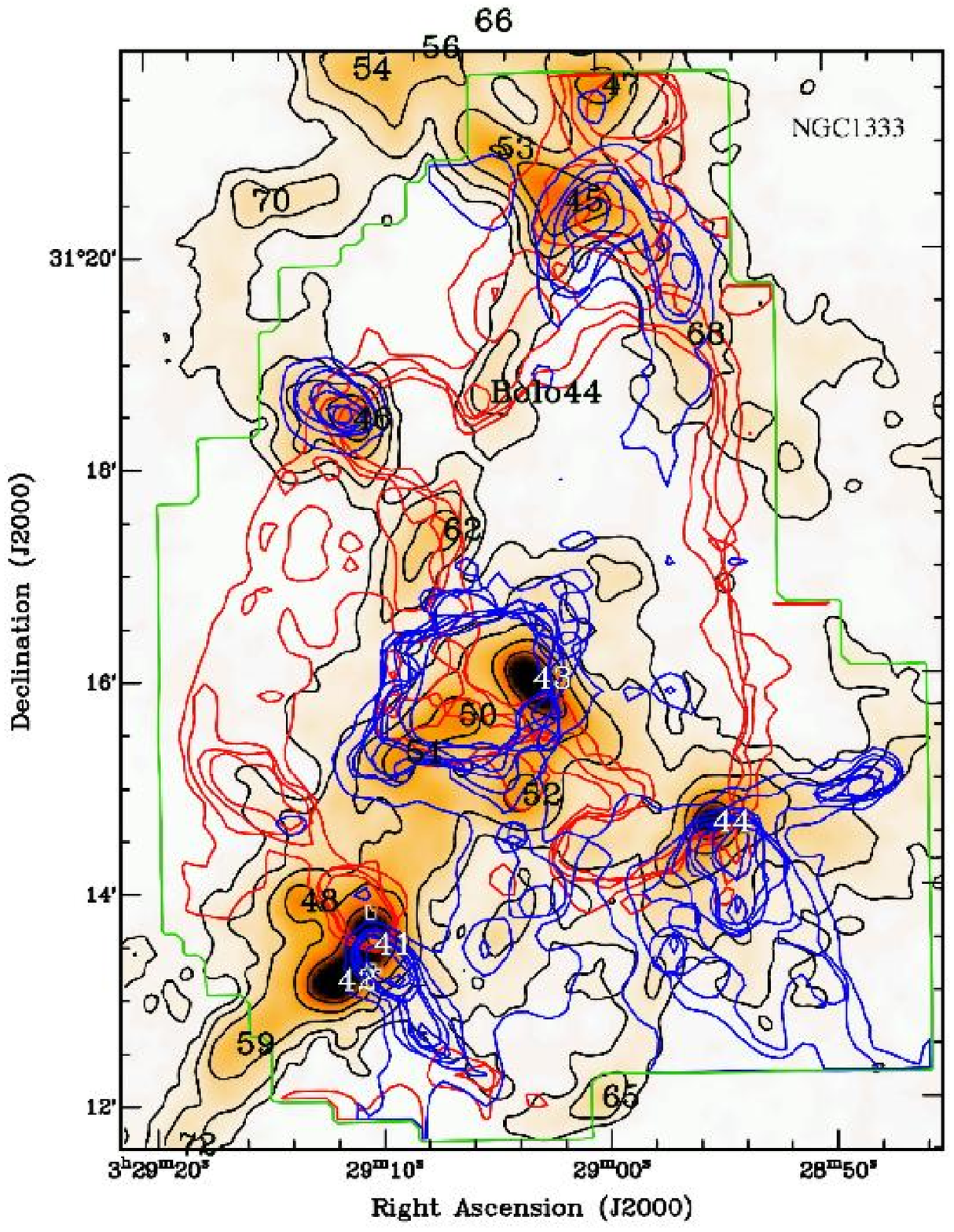}\\
\caption{continued. {\bf NGC1333} from data first published in \citet{kneesandell00}.  The renzogram intensity level is 3~K~km~s$^{-1}$ and included channels from -5--3~km~s$^{-1}$ (blue) and 12--18~km~s$^{-1}$ (red).  850\micron\ contours and colourscale are as before.}
\end{figure*}

\section{Outflow activity}
\label{sect:results}


\begin{table*}
\caption{Sources with outflow observations and resulting momentum
 fluxes.  For each source we list the core number and name (see \citet{class} for positions and core properties), the outflow status; the core ambient velocity $v_\mathrm{LSR}$, derived from C$^{18}$O, the maximum radius used in the momentum flux calculation, $\theta_2$ ($\theta_1$ was set to zero); and the resulting momentum flux and uncertainty.  Of the 51 outflow sources, 24 have definite outflows, 14 have no outflow detection, and 13 (9 with infrared detections) have possible but confused outflows.   }

\label{tbl:pflux}
\begin{tabular}{l l | l l |l c r}
Nr. &Name$^{1}$ &Outflow  &Class$^{2}$ &$v_{\mathrm{LSR}}$      &$\theta_2$ & $F_\mathrm{CO}$ \\
    &   &(y/n?)         &      &km~s$^{-1}$             & $''$          & $10^{-7}~M_\odot \hbox{km s}^{-1} \hbox{ yr}^{-1}$ \\
\hline

1      &b1-c            &y            &0 &6.4   &12.0 & 1.7   ( 0.8)\\
2      &b1-bS           &y            &0 &6.8    &30.0&12.3   ( 1.0)\\
4      &b1-d            &y            &0 &6.7    &15.0& 1.8   ( 0.3)\\
7      &IRAS 03301+3057 &y            &I &6.6    &30.0&12.6   ( 0.8)\\
10     &B1 SMM11        &y            &I &6.6    &30.0& 4.4   ( 0.4)\\
12     &HH211           &y            &0 &8.8    &30.0& 6.2   ( 0.8)\\
13     &IC348 MMS       &y            &0 &8.6    &30.0& 5.1   ( 0.5)\\
14     &                &y?           &I &9.1    &10.0& 1.2   ( 0.3)\\
15     &IC348 SMM3      &y            &0 &8.8    &30.0& 0.6   ( 0.2)\\
16     &                &n            &S &8.8    &--  &--     --    \\
17     &                &y?c13$^3$     &S &8.4   &10.0 & 0.8  ( 0.6)\\
18     &                &n            &S &8.6    &-- &--     --    \\
19     &                &n            &S &9.5    &-- &--     --    \\
20     &                &n            &S &8.5   &--   &--     --    \\
21     &                &n            &S &8.6   &--   &--     --    \\
23     &                &n            &S &9.0   &--   &--     --    \\
24     &                &n            &S &8.7   &--   &--     --    \\
25     &                &n            &S &9.2   &--   &--     --    \\
26     &                &n            &S &8.8   &--   &--     --    \\
27     &L1448 NW        &y            &0 &4.4    &30.0&27.5   ( 3.0)\\
28     &L1448 N A/B     &y            &0 &4.4    &30.0&36.4   ( 2.9)\\
29     &L1448 C         &y            &0 &4.8    &30.0&20.9   ( 1.5)\\
30     &L1448 IRS2      &y            &0 &4.1    &30.0&10.1   ( 0.8)\\
31     &                &y?c30$^3$     &0 &4.1   &15.0 & 3.6  ( 0.5)\\
35     &L1455 FIR4      &y            &I &5.3    &30.0& 9.4   ( 0.6)\\
36     &                &y            &0 &5.8    &20.0& 6.6   ( 0.5)\\
37     &L1455 PP9       &y?           &I &5.3    &30.0& 3.2   ( 0.4)\\
39     &L1455 FIR1/2    &y?           &I &5.3    &20.0& 1.8   ( 0.2)\\
40     &                &y?c35,36$^3$  &S &5.4     &30.0 & 2.6  ( 0.3)\\
41     &NGC1333 IRAS4A  &y            &0 &7.6    &30.0&45.5   ( 2.9)\\
42     &NGC1333 IRAS4B  &y            &0 &7.5    &10.0&10.4   ( 1.6)\\
43     &NGC1333 SVS13   &y            &I &7.9    &30.0&195.9  (14.2)\\
44     &NGC1333 IRAS2A  &y            &0 &7.7    &30.0&72.7   ( 4.4)\\
45     &NGC1333 SK24 &y            &I &7.6    &30.0&50.3   ( 3.4)\\
46     &NGC1333 SK20/21 &y            &0 &7.9    &30.0&59.5   ( 3.3)\\
47     &NGC1333 SK31    &y            &0 &7.6    &30.0&19.2   ( 1.9)\\
48     &NGC1333 IRAS 4C &y?c41,42$^3$  &0 &7.6  &20.0    & 2.7  ( 1.6)\\
50     &NGC1333 SK15 &y?c43,51$^3$  &I &7.9   &20.0 &166.2 (10.9)\\
51     &NGC1333 SK16    &y?c43,50$^3$  &S &7.6   &20.0 &54.5  ( 8.2)\\
52     &NGC1333 SK14 &y?c43,50$^3$  &0 &7.6   &30.0 &32.5  ( 2.6)\\
53     &NGC1333 SK26    &y?c45$^{3,4}$ &S &7.6   &20.0 & 4.4  ( 0.6)\\
61     &                &n            &0 &7.6   &--   &--     --    \\
62     &NGC1333 SK18    &y?c46$^3$     &0 &7.9   &20.0 &25.6  ( 4.1)\\
64     &NGC1333 Per 4A3/4D &n            &S &7.4   &--   &--     --    \\
68     &NGC1333 MBO146   &y?c45$^3$     &0 &7.7   &30.0 &27.5  ( 2.3)\\
76     &IRAS 03292+3039  &y            &0 &6.9    &10.0& 1.2   ( 0.5)\\
77     &IRAS 03282+3035  &y            &0 &7.2    &30.0& 4.8   ( 0.5)\\
78     &B5 IRS1          &y            &I &10.0    &10.0& 1.5   ( 0.2)\\
79     &                 &n            &S &10.0    &--&--     --    \\
80     &IRAS 03235+3004  &n            &I &5.0    &--& -- --\\
81     &B1 SMM1          &n            &I &5.9    &--& -- --\\

\end{tabular}

$^1$ Common name.  For positions, further names, references and further data see \citetalias{class}.

$^2$ From SED-based evolutionary indicators \citepalias{class}.

$^3$ Sources of potential confusion.

$^4$ At edge of mapped area.  Possible S blue lobe.

%

\end{table*}


\begin{figure*}[t]
\centering
\includegraphics[scale=1.0,angle=-90]{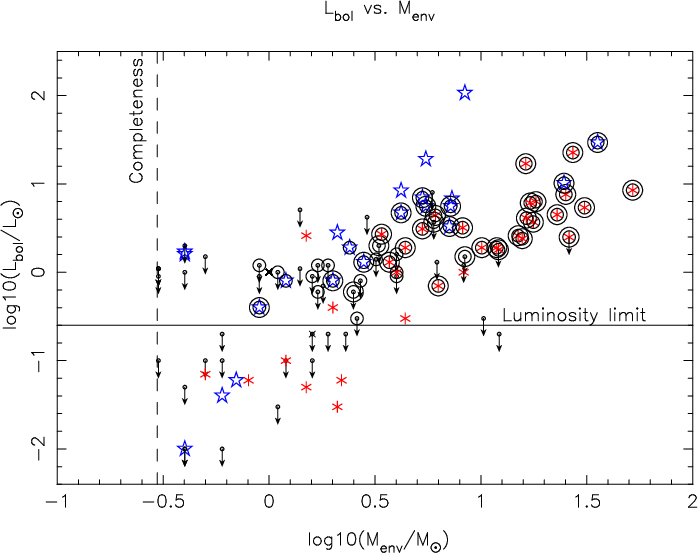}\\
\caption{\Lbol\ vs. \Menv\ for all submm cores in Perseus
  \citepalias{paperI,class} with $^{12}$CO targets marked by small
  black circles and outflow detections additionally marked by large
  black circles.  Class~0 sources are marked as asterisks ($\ast$,
  red) and Class~I by open stars (blue).  The remaining sources (dots)
  are assumed starless.  The horizontal line at 0.25~\Lsun\ marks the
  lower luminosity limit for the outflow sample, and the vertical
  dashed line at 0.3~\msun\ marks the $3\sigma$ detection limit for
  the SCUBA survey.}
\label{fig:sample}
\end{figure*}

Maps of the outflows superimposed on the submm continuum map are shown
in Fig.~\ref{fig:outflows}.  We identified sources with outflows based
on the objective criterion of $^{12}$CO linewings more than 1.5~K
($3\sigma$) at 3~km~s$^{-1}$ from the line centre (as measured in
C$^{18}$O).  Of the 51 sources in Table~\ref{tbl:pflux} observed in
$^{12}$CO~3--2, 37 sources (73\%) were classified as outflows on this
basis.  Spectra for these sources at the source position and the
outflow lobes are shown in Fig.~\ref{fig:outflowspec} (online version
only).  In some cases the outflow lobe spectra are missing due to
confusion or because the driving source lies at the edge of the mapped
area.  From the maps (Fig.~\ref{fig:outflows}), we identify 13 cases
where the broad linewings may be wholly or partly due to nearby flows
driven by other sources in crowded regions.  These uncertain flows are
marked with `y?' in Table~\ref{tbl:pflux}.  The remaining 14 sources
show no evidence for outflow activity in the form of either line wings
or red/blueshifted lobes, as shown in Fig.~\ref{fig:nooutflowspec}
(online version only).

All but four of the sources with outflows on the basis of linewings
also have a near-infrared detection from Spitzer IRAC
\citepalias{class}, \citep{jorgensen07}.  The exceptions are 17, 40,
51 (SK16) and 53 (SK26), all of which are classified as starless
(Table~\ref{tbl:pflux}).  In all of these cases the linewings are
likely due to confusion.  In the following analysis of the outflow
properties, we leave out these four sources, assuming that they are
indeed starless.  For most of the protostars it is the driving source which
is being detected by Spitzer but in some cases the infrared emission
may be due to HH objects (examples are HH211 (12), and (15)).  In 9 of
the 13 uncertain outflow sources, their protostellar nature is
confirmed by IR detections.  All but one of the sources with Spitzer
IRAC detections is observed in CO have molecular outflows.  The
exception is in NGC1333: the Class~0 source 61 lies to the north of
the luminous Herbig~Ae-Be star NGC1333~IRAS~8/SSV3 and B star
IRAS~9/BD$+30^\circ\,549$ and the interaction with radiation and winds
from these sources may well explain the outflow non-detection.  The
detection of outflows from all but one Spitzer-confirmed protostar
shows that both types of observation are similarly effective in
identifying protostars among submm cores.  Outflow-based
identification avoids the Spitzer confusion with extragalactic objects
but can be difficult in crowded regions where the outflows overlap.

\subsection{Outflow detection statistics}
\label{sect:detection}

\begin{figure}[t]
\centering
\includegraphics[scale=0.90,angle=0]{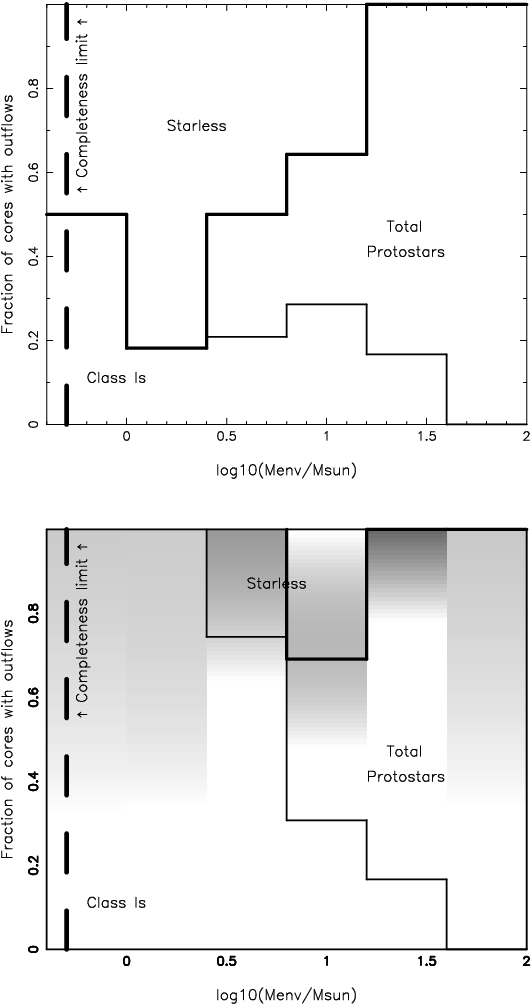}

\caption{{\bf Top:} Fraction of outflow sources as a function of core mass for sources with definite outflows (solid thick line).  The thin line shows the number of Class I sources as classified in \citetalias{class}.  The ratio of protostars to starless cores decreases with decreasing mass.  {\bf Bottom:} As above with the fraction of protostars corrected for the outflow detection statistics in that mass bin.  The greyscale gives the probability distribution of the fraction of protostars (see Appendix).}

\label{fig:outflowstats}
\end{figure}

\begin{figure}[t!]
\centering
\includegraphics[scale=0.8,angle=0]{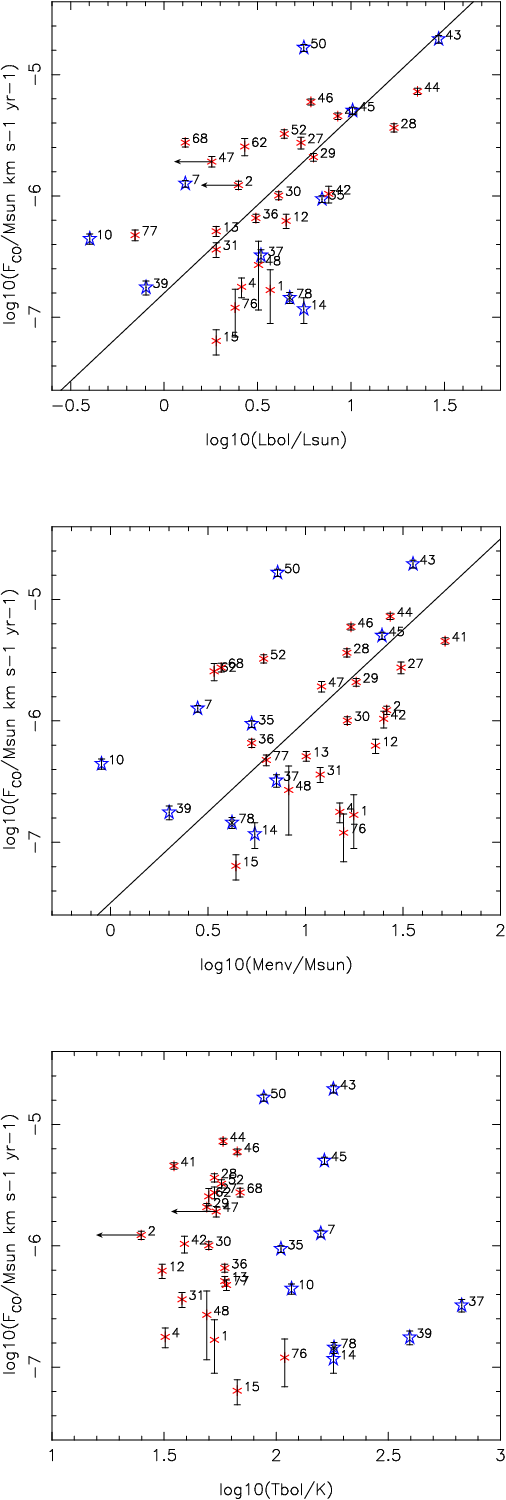}\\

\caption{Relationships between momentum flux and (top to bottom) bolometric luminosity \Lbol,
  core mass \Menv, and bolometric temperature \Tbol.  Sources marked by open stars (blue) and asterisks (red) are Class~I and Class~0 respectively based on the SED \citepalias{class}.  The top two panels show best fits corresponding to $F_\mathrm{CO} \propto \Lbol^{1.5}$ and $F_\mathrm{CO} \propto \Menv^{1.5}$ respectively.  Uncertainties on $F_\mathrm{CO}$ (see Sect.~\ref{sect:pflux}) are shown; uncertainties on \Lbol\ are typically 30\%, \Menv\ at least a factor 2, and \Tbol\ a few Kelvin.}
\label{fig:pflux}
\end{figure}

If we consider the outflows as a function of core mass
(Fig.~\ref{fig:outflowstats}, top), we find that the fraction of
sources with outflows decreases with decreasing core mass.  In
order to estimate our ability to detect outflows in the lower mass
bins, we performed the following simulation.  We considered the 14
outflow sources in the three highest mass bins ($\Menv > 1$~\msun).
We then simulated outflows from lower mass cores by scaling the
outflow spectra (Fig.~\ref{fig:outflowspec}, online version only) for
these 15 sources according to the modelled core mass as
$\hbox{(intensity/original intensity)} = \hbox{(model mass/original
  mass)}^{1.5}$.  The motivation for this power law scaling comes from
Fig.\ref{fig:pflux} which shows a rough correlation between momentum
flux and core mass with a gradient of $1.5\pm0.5$ (see Sect.~\ref{sect:pflux}).  We also formed
the simulated outflow maps by multiplying the original $(x,y,v)$
datacube by these scaling factors and adding back on the unscaled
noise levels (from sampling the noise in the outer velocity channels
of the unscaled maps).  Several scaling factors were used,
corresponding to moving the core masses an integer number of mass bins
to the left.  By this means we generated an artificial population of
spectra and maps of outflows for cores in the range 0.6--50\msun.  We
tested the spectra according to our criterion for linewings ($>1.5$~K
at 3~km~s$^{-1}$ from line centre).  We also additionally displayed
these maps using our usual visualisation software kvis \citep{gooch96} and
considered whether we could visually detect the resulting outflows
above the noise.  In all cases where the linewings satisfied the
objective criteria, we could see the outflows in the maps.  The
detection statistics for our simulated population are given in
Table~\ref{tbl:outflowstats}.

\begin{table}[h]
\begin{tabular}{l |l l l l l l}
$\log_{10}(M/\msun)$            &-0.2 &0.2 &0.6 &1.0 &1.4 &1.8 \\
\hline
Simulated maps              &14 &14  &14  &14  &11  &1\\
Sim. detections          &0 &1   &4  &13  &11  &1\\ 
Sim. rate          &0\%  &7\% &29\% &93\% &100\% &100\%\\
Obs. rate          &50\%  &18\% &50\% &64\% &100\% &100\%\\
\end{tabular}
\caption{Fraction of simulated outflows detected in each mass bin, compared to observed detection fraction.}
\label{tbl:outflowstats}
\end{table}

There is a decline in our detection ability at lower mass, and in the
lowest mass bins we detect few of the model outflows.  In the lowest
two mass ranges in Fig.~\ref{fig:outflowstats}, 0.4--2.5~\msun, 0 and
1 respectively out of 14 of the simulated low-mass outflows were
detected.  In the next highest mass bin (2.5--6~\msun), the detection
rate improved and 29\% of simulated outflows were detected.  The mean
detection rate below 6~\msun is predicted to be 12\%, or roughly 1 in
8 outflows detected.  

The outflow detection rate predicted by the simulation is lower than
the actual detection rate for all sources below 6~\msun.  This can be
seen in the lowest three mass bins of Table~\ref{tbl:outflowstats}.
This is true even if we include all the submm cores from the SCUBA
survey, and not just those with existing outflow observations.  We
have already detected 14 outflows among 75 submm cores with masses
below 6~\msun\ (Fig.~\ref{fig:sample}), or 19\% compared to an average
of 12\% for the simulated flows.  This suggests that our assumptions
about how outflow linewings scale with core mass are pessimistic,
and we are more likely to detect flows than the simulation predicts --
due to higher optical depth in more massive flows, for example.

The bottom panel of Fig.~\ref{fig:outflowstats} shows our best
estimate of the true fraction of outflow sources in each mass bin,
corrected for the detection statistics using (true outflow fraction) =
(fraction of outflows detected) / (detection probability in this mass
bin).  The calculation of the uncertainties (greyscales) shown in the
figure is given in Appendix~\ref{app:outflowprob}.  In mass bins where
the observed detection rate equals or exceeds the simulated rate, this
model would predict that all the apparently starless cores contain
undetected protostars.  Only in the 6--16~\msun mass range is the
predicted number of outflow sources smaller than the number of
apparently starless cores.  If this model is correct, then we fail to
detect outflows from the majority of protostars at masses $<6 \msun$.

Thus, the modelling suggests that the low number of outflow detections from low mass cores  is a selection effect.  




If this is the case then this outflow survey and Spitzer c2d
\citet{c2d} must have very similar sensitivities to embedded
protostars as only one mid-IR detected sources in the sample had no
corresponding outflow and conversely only four outflow sources
had no mid-IR detection.  Therefore any selection effect must operate
both in the infrared and molecular lines.  Orientation could help
produce such close agreement as infrared sources are hard to detect
where disks are edge-on and outflows are hard to detect when they lie
close to the plane of the sky.  Assuming random outflow orientation,
though, the number of infrared non-detections due to edge-on disks
(within $10^\circ$ of the line-of-sight) can only be $\sim 20\%$ of
the total protostellar fraction.

This modelling suggests that the low-mass cores without outflow or
Spitzer detections may also be protostellar, but of course does not prove
that they are.  For one thing, the assumptions in our model about how
outflow masses vary with core mass may not be correct.  Only
further searching for outflows, infrared or radio sources towards the
apparently starless cores can determine the true nature of the
apparently starless cores.  Note that such observations would need to
be with higher sensitivity than this outflow survey, the Spitzer Cores
to Disks programme \citep{c2d}, or its successor, the Spitzer Gould's
Belt survey, in order to identify less luminous protostars.

\section{Momentum fluxes}
\label{sect:pflux}

Outflow momentum fluxes (or outflow luminosities) are given in
Table~\ref{tbl:pflux} for all the possible outflow sources.  The
method is developed from the momentum flux per beam calculation of
\citet{fullerladd02}, rather than calculating the momentum flux within
an annulus \citep{batc96}, though the two methods should yield
comparable results.  The momentum flux crossing a cylinder radius
$\theta_r$ perpendicular to the plane of the sky and centred on the
source, contributed by the mass within one beam, is calculated
individually for each beam as:

$$ \dot P = \sum_{v_\mathrm{obs}}{M_\mathrm{B}(v_\mathrm{obs})\, v_\mathrm{obs} / \cos i \over D\theta_\mathrm{B} / v_\mathrm{obs} \tan i}$$

where $v_\mathrm{obs}= |v_i - v_\mathrm{LSR}|$ is the observed velocity
offset from the core systemic velocity $v_\mathrm{LSR}$ (determined
from C$^{18}$O observations of the same region at $1'$ resolution
\citepalias{paperI}) , $i$ is the angle the
outflow makes with the line-of-sight, $D$ the source distance and
$\theta_\mathrm{B}$ the beam FWHM.  The numerator gives the momentum
in the beam ($v = v_\mathrm{obs}/\cos i$ is the outflow radial
velocity) and the denominator the time taken to cross the beamwidth at
velocity $v$.  The sum is over velocity channels.

The mass within one beam $M_\mathrm{B}$ is calculated from the CO
integrated intensity over the velocity channel $I_\mathrm{CO}\,\Delta
v$ assuming a CO abundance of $X_\mathrm{CO} = 10^{-4}$
\citep{frerking82,wilsonrood94} and a LTE excitation at an outflow
temperature of 50~K resulting in $N_{\mathrm{H}_2} = 2.5\times 10^{19}
(\int T_\mathrm{MB}^* \, dv / \hbox{K km s}^{-1}) \hbox{ cm}^{-2}$.  A
temperature of 50~K is chosen as there is increasing evidence that
outflows contain warm gas
\citep{hatchell99a,hatchell99b,nisini00,giannini01} but the conversion
  is relatively insensitive to the temperature, increasing by only
  40\% over the temperature range 20--100~K.  The mass within one beam
  is calculated as $M_\mathrm{B}=(\pi/4)\, D^2 \theta_\mathrm{B}^2
  N_{\mathrm{H}_2} m_{\mathrm{H}_2}.$

The momentum flux contributions from individual beams are then
summed in annuli, according to:

\begin{eqnarray}
F_\mathrm{CO} &=& \sum_\mathrm{annulus}{2\pi\theta_r\over N_\mathrm{pix} \theta_B} \dot P\nonumber\\
              &=& \sum_{\theta_1 < \theta_r < \theta_2} {{\pi^2\theta_r D \over 2N_\mathrm{pix}} \sum_{v_\mathrm{obs}} N_{\mathrm{H}_2}(v_\mathrm{obs})\, m_{\mathrm{H}_2} v_\mathrm{obs}^2 {\sin i\over\cos^2 i}}\nonumber
\end{eqnarray}

where $\theta_r$ is the angular distance of each pixel from
the outflow centre, $D$ is the source distance, and the average is
over pixels between radii $\theta_1$ and $\theta_2$ with
$N_\mathrm{pix}$ the number of pixels between those radii.  As the
contribution to the momentum flux is calculated individually for each
beam, and varies little between neighbouring annuli, we set the inner radius to
zero.  The factor $2\pi\theta_r/N \theta_B$ corrects for the
overcounting because the $14''$ beam size is larger than the $5''$ pixel size.

The driving source positions were taken to be the SCUBA emission peaks
from Paper~I. To reduce the contribution due to emission outside the
outflow region, the outflow areas to be included in the momentum flux
calculation were selected by masking with $^{12}$CO integrated
intensity at a cutoff of 3 times the noise level in the integrated
intensity map.    We default to calculating the momentum
flux within a circle of outer radius $30''$ unless it was clear from
the map that the outflow was shorter than this or there was potential
confusion within this radius, in which case the radius of the circle
was reduced.  In practice we find the choice of circle or annulus and
the radius of the circle (except for the case of confusing flows)
makes little difference to the resulting momentum fluxes, as expected
at radii where there is no large variation in the injection of
momentum.  

The inner velocity channels, contaminated by ambient emission, were
cut out to $\pm~2$~km~s$^{-1}$ and the highest velocity considered was
$\pm~15$~km/s.  Ideally, for a fair comparison of the momentum flux
between strong (massive) and weak flows, one wants good
signal-to-noise (S/N) in the linewings out to similar velocities.
With roughly constant sensitivity and line strengths varying by an
order of magnitude the strong linewings can be measured with good S/N
to higher velocities in the stronger flows.  Adjusting the velocity
range on a flow-by-flow basis according to where the S/N is good
therefore introduces a strong bias in momentum flux towards massive
flows.  Not only are these more massive but also more of the high
velocity channels are included in the calculation.  Therefore we
consider it fairer to use the same velocity cut for each source, even
though this excludes the contribution from the highest velocity
channels in all flows.

The uncertainties on the momentum flux, listed in
Table~\ref{tbl:pflux}, are the root sum-of-squares of the
uncertainties of each contributing pixel.  The pixel uncertainty is
taken to be the difference from the mean momentum flux in the
contributing (integrated intensity $> 3\sigma$) region.

We have not applied a correction factor for either inclination or
optical depth, unlike \citet{batc96}.  Therefore our momentum fluxes
are strictly lower limits.  \citet{batc96} point out that inclination
$i$ reduces the momentum flux by a factor $\sin i / \cos^2 i = 2.9$
for a mean angle of 57.3 degrees.  For an optical depth $\tau$,
uncorrected CO column densities are underestimated by a factor $\tau
/(1-\exp(-\tau))$ or a typical factor of 3.5 so that true momentum
fluxes are typically a factor of 10 greater than measured.  Outflow
optical depths are high even in the $J=3\hbox{--}2$ line with with
$\tau_{3-2} > 4$ in the line wings \citep{margulis85,hogerheijde98}.


The (uncorrected) momentum flux results for all the possible outflow
sources are given in Table~\ref{tbl:pflux}.  Comparing the sources in
common, our momentum fluxes are typically lower by a factor of a few
than previous analyses by \citet{batc96} (taking the optical depth /
inclination correction into account) and {kneesandell00}.  The main
reason for this is that we assume the outflows are warm (50K), which
reduces mass estimates by a factor of $\sim 4$ over the 10K assumed by
\citet{batc96}.  This is a systematic difference, reducing the
momentum fluxes for all sources by the same fraction.  For flows with
high velocity gas, we can also underestimate momentum fluxes because
we do not include the highest velocity channels (as discussed above).
For two sources, IRAS~03282+3035 (77) and B5~IRS1 (78), we measure
momentum fluxes more than a factor of 10 lower than previous
estimates.  We have no reason to suspect the calibration for these two
maps, and they do not have particularly broad linewings (as can be
seen in Fig.~\ref{fig:outflowspec}, online).  It is possible that in
these flows, the momentum flux measured at greater radii in the flow
is larger than in the region covered by our $2'$ maps.

\subsection{Momentum flux vs. \Lbol, \Menv\ and \Tbol}

The outflow momentum flux \Fco\ is plotted versus core mass \Menv,
bolometric luminosity \Lbol\ and bolometric temperature \Tbol\ in
Fig.~\ref{fig:pflux} \citepalias[see ][ for the calculation of those
quantities from the SEDs]{class}.  The well known relationship between
momentum flux and bolometric luminosity \citep{batc96,cabritbertout92}
is evident despite the intrinsic scatter in the momentum flux due to
uncorrected outflow inclination.  This correlation is often taken as
evidence that the outflow and the accretion luminosity are both being
powered by the same mechanism, infall, and despite the details of the
magnetic ejection/accretion process in the inner few tens of AU of the
disk, the energy input into each remains strongly linked.  The
correlation is tighter towards higher \Lbol\ and holds for both
Class~0 and Class~I sources (marked by red asterisks and blue open
stars, respectively; see \citetalias{class} for details of classification).

\begin{figure}[b!]
\centering
\includegraphics[scale=0.8,angle=0]{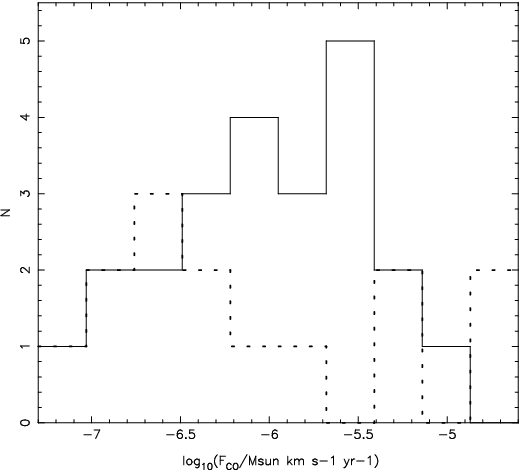}\\

\caption{Histograms of momentum flux for Class~0 (solid) and Class~I (dotted) protostars.}
\label{fig:pfluxstats}
\end{figure}

\citet{batc96} found a decline of momentum flux between the Class~0
and Class~I phases which has been taken as evidence that outflow power
declines with age.  In our sample, this is less clear cut.  The mean
$\log_{10}(\Fco/\msun \hbox{km s}^{-1} \hbox {yr}^{-1})$ of the
Class~0 population is $-6.0$ with a sample standard deviation of
$0.6$, close to the Class~I mean of $-6.1$ (standard deviation
0.8).  A histogram of the momentum flux distribution for Class~0 and
Class~I sources (Fig.~\ref{fig:pfluxstats}) shows that both of the
distributions are very flat. Applying the Kolmogorov-Smirnov (K-S)
test indicates a 27\% probability that the two samples come
from the same parent distribution and therefore there is no evidence
for evolution in momentum flux between Class~0 and Class~I.  There is
considerable scatter with objects of both classes with both low\ and
high-\Fco.

In particular, three Class~I objects-- NGC1333~SVS13 (43), NGC1333
ASR114~(45) and NGC1333 HH7-11 MMS4 (50)-- have particularly high
momentum flux and four Class~0 sources -- b1-c (1), b1-d (4), B1~SMM5
(76) and IC348~SMM3 (15) -- have low momentum flux.  The third panel of
Fig.\ref{fig:pflux}, in which momentum flux is plotted against \Tbol,
shows clearly that the Class~Is (on the right hand side of the plot)
have more scatter in momentum flux than the Class~0s.  The two high
\Tbol\ sources are the well-known Class~I sources L1455~PP9 (37),
which may contain multiple protostars, and L1455~FIR1/2 (39).

All three of the high momentum flux Class~I cores, NGC1333~SSV13 (43),
NGC1333~ASR114 (45) and HH7-11 MMS4 (50) are in the confused NGC1333
region.  HH7-11 MMS4 (50) is a confused source where a major
contribution to the momentum flux comes from the HH7-11 driving source
SSV13.  Both SSV13 and ASR114 contain multiple protostars as evidenced
by radio emission \citep{rac97,kneesandell00}.  Can this explain their
high momentum fluxes despite Class~I classification?  NGC1333~SSV13 (43) is
a multiple source (SSV13 and 13B) potentially consisting of a Class~I
source (providing the infrared-based classification) and a Class~0
source (powering the outflow).  The flows from SSV13/13B are confused
and the momentum flux includes contributions from both, though
\citet{kneesandell00} attributed more momentum flux to the Class~I
SSV13 than the potential Class~0 counterpart SSV13B.  Likewise ASR114
(45) contains multiple embedded radio sources \citep{rac97}, and the
outflows are confused by the flow from the $4'$ distant SSV13B which
is believed to drive the nearby HH~12 \citep{kneesandell00}, and the
momentum fluxes could be overestimated by the contribution from the
flow from the south.  It is possible that in both these cases the high
momentum flux is due to a Class~0 source, either in the same submm
core or nearby -- an open question which could be resolved by
interferometric observations of the flows.


However, the Class~0 sources with low momentum flux -- b1-c (1), b1-d
(4), IC348-SMM3 (15) and B1-SMM5 (76) -- are less easy to explain
away.  Three of these sources (1, 4, 76) also have low bolometric
luminosity for their mass.  Our sample, because of its submm-based
sample selection, may well contain a population of low-luminosity,
low-\Fco\ Class~0 sources which were not included in earlier studies.



There are physical reasons why outflow force may vary between sources
at the same evolutionary stage.  The momentum in the driving wind
depends on the mass accretion rate and in turn on the history of the
mass reservoir available for accretion.  For example, this may differ
between clustered and more isolated protostars \citep{henriksen97},
though this does not appear to explain the results in Perseus.  Early
Class~0 sources may not have reached their full accretion rates
\citep{smith00}.  The low momentum flux Class~0 sources b1-c (1), b1-d
(4) and B1-SMM5 (76) which also have low bolometric luminosity are
likely to be early Class~0 sources which have not yet reached their
full accretion power.  The fraction of the wind momentum which is
transferred to the molecular outflow depends on entrainment of the
ambient material, which may vary in efficiency.  In the competitive
accretion model, mass may continue to flow onto already-accreting
envelopes and boost accretion at late stages, which would lead to a
natural scatter in accretion rates.  Jet ejection is known to be
episodic \citep{raga90,raga04} but the swept-up molecular outflow
smooths out any variability on timescales of $\sim100$ years, such as
those which produce jet knots.

\begin{figure}[t!]
\centering
\includegraphics[scale=0.8,angle=0]{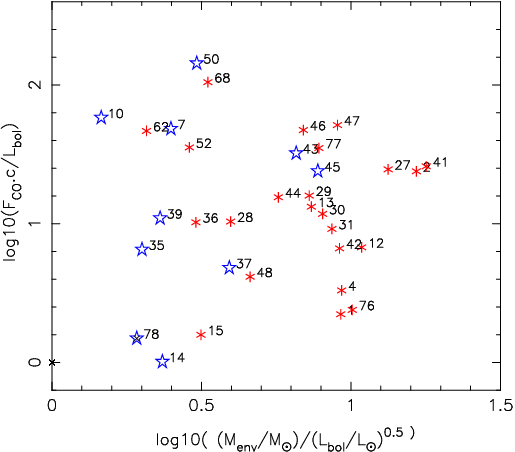}\\

\caption{$F_\mathrm{CO}\, c / \Lbol$ (dimensionless) vs. $\Menv/\Lbol^{0.5}$ (in $\msun$ nd $\Lsun$ respectively). \citet[See also][ Fig.~7]{batc96}.  Sources marked by open stars (blue) and asterisks (red) are Class~I and Class~0 respectively based on the SED \citepalias{class}.}
\label{fig:lumcorr}
\end{figure}

\citet{batc96} show in their Fig.~4 a plot in which momentum flux and mass are compensated for source luminosity.  In Fig.~\ref{fig:lumcorr} we show the analogous figure for our dataset.  For our sources the relationship between mass and luminosity is $\Lbol \propto \Menv^{1.96} $ \citepalias{class} so the abscissa becomes $\Menv/\Lbol^{0.5}$ and the ordinate  $F_\mathrm{CO}\, c / \Lbol$.  The luminosity correction separates the Class~I sources from the Class~0s, which tend to be more massive for the same luminosity.  This separation in mass is seen in Fig.~\ref{fig:lumcorr}, with the Class~0 sources lying to the right of the diagram.  It also takes out the effect of distance, though this is irrelevant for our Perseus-only sample.

The luminosity correction has to be treated with caution because it does not differentiate between luminosity and momentum flux evolution.  Evidence for luminosity evolution comes from \citetalias{class}, which shows that Class~0 sources have, on average, lower luminosities at a given mass than Class~I sources.  Some evolutionary schemes \citep{smith00} interpret this as an increase in luminosity between the Class~0 to Class~I phases.  On the luminosity-corrected diagram, decreasing luminosity can be mistaken for increasing momentum flux.

Nonetheless, in the luminosity-corrected diagram, the $F_\mathrm{CO}\, c / \Lbol$ values for the Class~0 sources are similar to their Class~I counterparts, with no obvious evolution between the two stages.

There is a general trend of increasing momentum flux with \Menv\ 
(middle panel of Fig.\ref{fig:pflux}) although there is more than an
order of magnitude of scatter, as found for \Fco--\Lbol.  A physical
explanation for such scatter is that the energy in the mass infall,
which powers both the mass outflow and the bolometric luminosity, is
dependent on the infall rate which is not set by the current mass of
the core but by its past history.  \citet{hogerheijde98} found a
better correlation of \Fco\ with \Menv\ than \Lbol, but we see no
evidence for this.


In summary, there is broad agreement between our results on the
momentum flux and previous studies, with similar scatter on the
correlations with \Lbol\ or \Menv\ despite the consistent derivation
of these quantities and the single distance of our sample. A
difference is the significant population of low-\Fco\ Class~0 sources
and the small number of high-\Fco\ Class~Is which suggest that
evolution of \Fco\ from Class~0 to Class~I is not a simple decline but
that there may be variation within each class, eg. as luminosity and
outflow power increase at the beginning of the Class~0 phase.

\section{Mass loss rates and core destruction timescales}
\label{sect:lifetimes}

\begin{figure*}[t]
\centering
\includegraphics[scale=0.90,angle=-90]{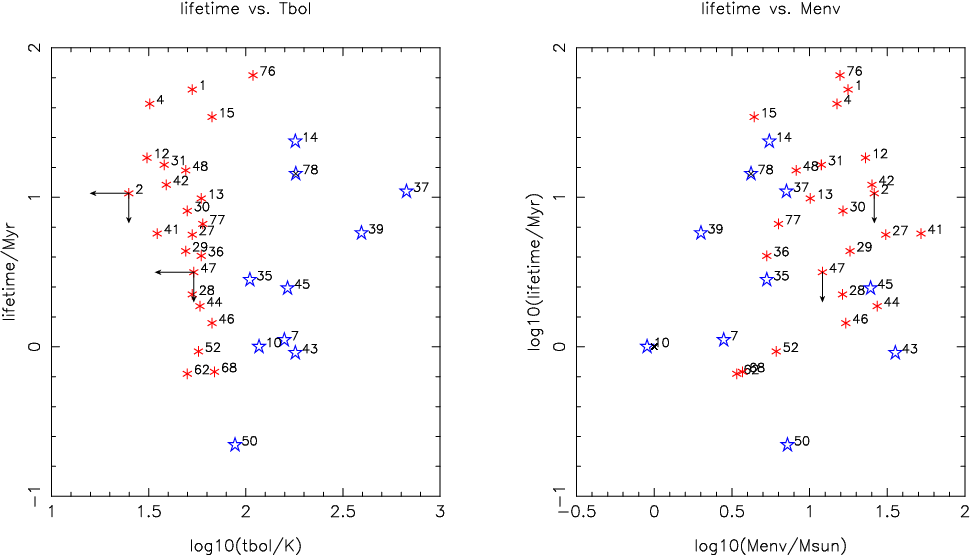}\\
\caption{Core destruction timescales at current outflow mass loss rate, vs.
  {\bf left} bolometric temperature; {\bf right} current core mass.  Class~I sources \citep[based on the SED]{class}, are marked by (blue) open stars and Class~0 sources by (red) asterisks.}
\label{fig:lifetimes}
\end{figure*}

A significant proportion of the momentum from the
driving jet/wind is deposited in the protostellar core and removes
matter from the core in the molecular outflow
\citep{lada85,parkerpadman91,fullerladd02,arcesargent06}.  The time it
would take to clear a protostellar core of mass via the outflow
is a constraint on the lifetime of the embedded protostellar phase,
especially when taken together with an indication of whether the
source is at the start (early Class~0) or end (Class~I) of its
embedded lifetime.

In addition to the outflow, mass is lost from the protostellar core by
accretion onto the disk/star.  The ratio of mass ejected in the
wind/jet to mass accretion, $f$, varies significantly with different
models for jet/wind launching.  Observations suggest that $f \simeq
0.3$ is a good estimate \citep{richer00}.  Although the mass loss in
the jet is then less than half of the total accretion, momentum
transfer to the slower molecular outflow can lead to mass loss rates
in the outflow that are a factor of 10 higher than the total mass
accretion rate: $\dot M_{CO} = f \dot M_\mathrm{acc}
(v_\mathrm{w}/v_\mathrm{CO}) \simeq 10 \dot M_\mathrm{acc}$, taking
the jet ejection fraction $f_a = 0.3$ and typical values for the jet
and CO velocities of 150 and 5~km~s$^{-1}$ respectively,
  and assuming all the wind/jet momentum is deposited in the cloud.
  At early times when this is true, before the outflow has broken out
  of the cloud, then mass loss in the outflow dominates by an
order of magnitude over the mass accreted onto the central object, and
can be taken as a good estimate of the total mass loss rate.  For more evolved flows less of the wind/jet momentum may be transferred to the CO and in the most evolved sources the accretion rate may begin to dominate over the mass loss in the outflow. 

The outflow mass loss rate \Mdot\ can be calculated from the momentum
flux \Pdot\ by dividing by the mean outflow velocity $\bar v$.  $\bar v$ is
a difficult quantity to estimate accurately.  At low velocities it is
difficult to separate outflow material from protostellar core.  Any
mass at high velocities can also make a large difference to $\bar v$ with
correspondingly high uncertainties.  Making an estimate using $\bar v =
\sum_i I_\mathrm{CO} |v_i-v_{\mathrm{LSR}}| / \sum_i I_\mathrm{CO}$ where $I_\mathrm{CO}$ is the
integrated spectrum over the observed inner section of the outflow, we
calculate values for individual flows which range from
1--10~km~s$^{-1}$ depending on where the integration limits -- inner
and outer velocity channel cutoffs -- are set.  We take a value of
$5\hbox{ km s}^{-1}$ as typical, in agreement with previous estimates
of $P$ and $M$ \citep{zhang05,kneesandell00}, with an uncertainty
of a factor of 2.

The resulting mass loss rates in the outflow range from $1\times
10^{-7}\hbox{--} 6\times 10^{-5}~M_\odot~\hbox{ yr}^{-1}$, from the
momentum fluxes in Table~\ref{tbl:pflux}, and assuming a factor of 10
to take into account inclination and optical depth into account (see
Sect.~\ref{sect:pflux} and \citet{batc96}).  These core mass loss
rates are similar to those calculated by \citet{fullerladd02} from the
rarer CO isotopologues in molecular cores.  Mass loss rates of above
$10^{-4}~\msun~\hbox{yr}$ have also been found for massive flows
\citep{zhang05,beuther02,churchwell97} and it is not a surprise to
find rates approaching this in Perseus, which is forming stars up
to spectral type B.

%
%
%
%

The destruction timescale $t_\mathrm{d}$ of the core at the current
mass loss rate can be calculated by dividing the current core mass
by the total mass loss rate, including the optical depth/inclination
correction factor of 10.  These timescales are plotted in
Figs.~\ref{fig:lifetimes} and range from $0.2\hbox{--}60$~Myr, with a
mean $\bar t_{\mathrm d} = 11$~Myr.  These destruction timescales are
then lower limits to the lifetime remaining to each core, if outflows
are the only source of mass loss and outflows only become weaker over
time.  This is believed to be true for all but the youngest Class~0
sources.

The sources with the longest lifetimes are the sources with the
highest mass to momentum flux ratios.  As luminosity and momentum flux
are correlated (Fig.~\ref{fig:pflux}), these are also the sources with
high mass-to-luminosity ratios and cold temperatures, a signpost of an
early Class~0 source.  Indeed, we find that the Class~0 sources b1-c
(1), b1-d (4), IRAS~03292+3039 (76) and the newly identified protostar
IC348-SMM3 \citep[15, Paper~I, ][]{walawender06} have long lifetimes
($>10$~Myr).  In all cases this is due to low momentum flux for the
given core mass.  Sources with particularly short destruction
timescales $\sim 10^4$ years are IRAS 03301+3057 (7), B1 SMM11 (10)
and HH7-11~MMS (43).  These sources have little mass in their
cores and are believed to be Class~I sources near the end of the
embedded phase.  Thus the objects with the most extreme lifetimes are
Class~0 for the longest lifetimes and Class~I for the shortest
lifetimes.  The absence of short lifetimes for low \Tbol\ sources is
because only high \Menv\ sources can produce a low \Tbol, and these
tend to have a long lifetime.  Low-mass sources have lower optical
depth in their cores and higher \Tbol\ for the same evolutionary
state, as discussed in \citetalias{class}.

\begin{figure}[b!]
\centering
\includegraphics[scale=0.8,angle=0]{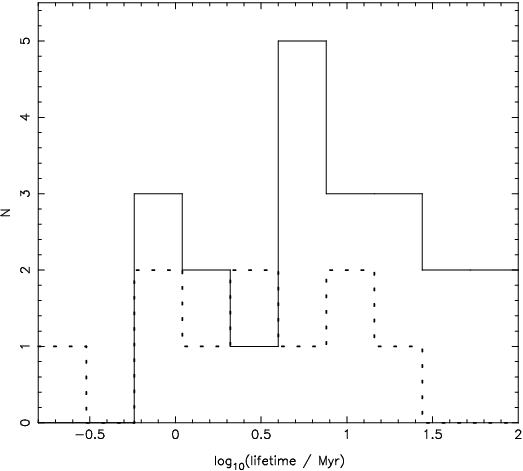}\\

\caption{Histograms of core destruction timescales for Class~0 (solid) and Class~I (dotted) protostars.  Sources with upper limits on the lifetime are excluded.}
\label{fig:lifetimestats}
\end{figure}

The mean destruction timescale for Class~Is (6
Myr) is shorter than for Class~0s (14 Myr), suggesting that there
might be some evolution between Class~0 and Class~I, with mass
declining faster than outflow force.  Yet the evidence for evolution
is weak: the scatter in $t_\mathrm{d}$ is large with both classes
showing a range of destruction timescales from a few times $10^5$ to
$\sim 10^8$ years at the current outflow rate, as shown in the
histogram of lifetimes (Fig.\ref{fig:lifetimestats}), and a K-S test
confirms a 43\% probability that these populations could be drawn
from the same underlying distribution.



%


The mean lifetime of the protostellar stage is estimated to be
$1.5\hbox{--}4\times 10^5$~years
\citep{greene94,kenyonhartmann95,wilking89,kenyon90,class}.  This is
significantly shorter than our mean $t_{\mathrm d}$ of $10^7$~years
for the removal of the core mass by outflows.  It was also
recently suggested by \citet{arcesargent06} that outflows may play an
important role in mass loss and core evolution during the Class~I
phase.  However, for the majority of sources, Class~0 and Class~I,
either we have overestimated the timescales due to systematic errors
or the bulk of the mass is removed by a means other than molecular
outflows.  This is compounded by the fact that not all of the mass
accelerated in the low velocity flow will have velocities sufficient
to remove it from the gravitational field of the central object, so
the outflows then give an upper limit on the mass loss rate to the
protostellar core.

There are several factors which could introduce systematic errors in
the calculation, which might result in an overestimate of the core
destruction timescales.  We may have overestimated the core
masses: as discussed in \citetalias{class}, our core masses lie at
the high end of the range consistent with the data, and could be
overestimated by a factor of 4 or more -- though half of this is due
to the uncertain distance, which affects dust and CO masses equally,
so again it is hard to see how this can explain more than a factor of
2 difference between protostellar lifetimes and core destruction rates
by outflows.  The outflow mass loss rates are calculated using several
uncertain factors in the conversion of CO~3--2 intensity to mass.
Firstly, there is the optical depth and inclination correction, for
which we assumed a factor of 10, and the outflow temperature, for
which we assumed 50~K, though this cannot affect the masses by more
than a factor of 1.5.  The mean velocity is also uncertain to a factor
$2$ but the destruction timescales would only reduce if it were lower
than the assumed 5~km~s$^{-1}$.

If protostellar cores are not destroyed by outflows and accretion,
where does the mass go?  A possibility is that we are detecting
material that will ultimately form not single stars but small
clusters, and stars will continue to form until the mass is used up,
as is seen in hydrodynamical models \citep[eg.][]{bbb03}.  Many of the
most massive cores are already known to contain multiple protostars
(Sect.\ref{sect:pflux},\citetalias{class}).  A further possibility is
that some cores will ultimately form massive stars and the molecular
cores will be destroyed by ionisation.  Finally, as mass is lost from
the system and the core heats up some material may become
gravitationally unbound and dissipate.

\section{Summary and conclusions}
\label{sect:summary}

We have searched for molecular outflows towards 51 of the 103 known
submm cores in Perseus listed in \citetalias{class} and based on the SCUBA
\citepalias{paperI} and
Bolocam \citep{enoch06} surveys.  The outflow sample is biased towards
the higher luminosity submm cores but the fraction of cores sampled is
fairly independent of mass except at the lowest masses.  The main results of this study are the following: 

\begin{enumerate}
\item We detected broad linewings from 37/51 sources surveyed. Of
  these, 13 cases could be due to confusion with neighbouring flows,
  though 9 are confirmed protostellar by Spitzer infrared detections.
  Only one source with a Spitzer detection has no apparent outflow.


\item The well-known correlation of momentum flux with bolometric
  luminosity is present, though with considerable scatter.  Momentum
  fluxes are not significantly higher for Class~0 sources than
  Class~I: on the basis of a K-S test, we cannot reject that the two
  samples are drawn from the same distribution.  There are three
  Class~I sources with apparently high momentum fluxes, possibly due
  to confusion with Class~0 flows, and there are several Class~0
  sources with low momentum fluxes, possibly because they are very
  young and have not yet reached full power.
  
\item Core destruction timescales based on constant outflow mass loss
  rates and core masses from submm emission range from $10^5$--$10^8$
  years, with no clear evolution between Class~0 and Class~I sources.
  These timescales are longer than the estimated lifetimes for
  protostars of $1.5\hbox{--}4\times 10^5$ years.  There are several
  possible explanations for this discrepancy. These include the
  possibility that outflows may not be the primary mechanism which
  destroys the cores. Alternatively, although it appears unlikely that
  the core mass is overestimated due to systematic issues by a large
  enough factor to account for the timescale discrepancy, it is
  possible that we are overestimating the fraction of the core mass
  bound to each protostar and that the material fragments forming
  multiple sources.

\end{enumerate}

\acknowledgements

Our thanks go to Jane Buckle, Ned Ladd and others who carried out the
$^{12}$CO observations, and to an anonymous referee for prompting
several improvements to the paper.  The James Clerk Maxwell Telescope
is operated by the Joint Astronomy Centre on behalf of the Particle
Physics and Astronomy Research Council of the United Kingdom, the
Netherlands Organisation for Scientific Research, and the National
Research Council of Canada.  JH acknowledges support from DFG SFB 494
and the PPARC Advanced Fellowship programme.  This research made use
of the SIMBAD query facility of the Centre de Donn\'ees Astronomiques
de Strasbourg and extensive use of the Karma visualisation package
developed by Richard Gooch at CSIRO \citep{gooch96}, which includes an
implementation of the Renzogram developed by Renzo Sancisi of the
Kapteyn Institute in Groningen.

\bibliographystyle{aa}
\bibliography{perseus}

\begin{thebibliography}{45}
\expandafter\ifx\csname natexlab\endcsname\relax\def\natexlab#1{#1}\fi

\bibitem[{{Arce} \& {Sargent}(2006)}]{arcesargent06}
{Arce}, H.~G. \& {Sargent}, A.~I. 2006, \apj, 646, 1070

\bibitem[{{Bate} {et~al.}(2003){Bate}, {Bonnell}, \& {Bromm}}]{bbb03}
{Bate}, M.~R., {Bonnell}, I.~A., \& {Bromm}, V. 2003, \mnras, 339, 577

\bibitem[{{Beuther} {et~al.}(2002){Beuther}, {Schilke}, {Sridharan}, {Menten},
  {Walmsley}, \& {Wyrowski}}]{beuther02}
{Beuther}, H., {Schilke}, P., {Sridharan}, T.~K., {et~al.} 2002, \aap, 383, 892

\bibitem[{{Bontemps} {et~al.}(1996){Bontemps}, {Andre}, {Terebey}, \&
  {Cabrit}}]{batc96}
{Bontemps}, S., {Andre}, P., {Terebey}, S., \& {Cabrit}, S. 1996, \aap, 311,
  858

\bibitem[{{Cabrit} \& {Bertout}(1992)}]{cabritbertout92}
{Cabrit}, S. \& {Bertout}, C. 1992, \aap, 261, 274

\bibitem[{{Churchwell}(1997)}]{churchwell97}
{Churchwell}, E. 1997, \apjl, 479, L59+

\bibitem[{{de Zeeuw} {et~al.}(1999){de Zeeuw}, {Hoogerwerf}, {de Bruijne},
  {Brown}, \& {Blaauw}}]{dezeeuw99}
{de Zeeuw}, P.~T., {Hoogerwerf}, R., {de Bruijne}, J.~H.~J., {Brown}, A.~G.~A.,
  \& {Blaauw}, A. 1999, \aj, 117, 354

\bibitem[{{Enoch} {et~al.}(2006){Enoch}, {Young}, {Glenn}, {Evans}, {Golwala},
  {Sargent}, {Harvey}, {Aguirre}, {Goldin}, {Haig}, {Huard}, {Lange},
  {Laurent}, {Maloney}, {Mauskopf}, {Rossinot}, \& {Sayers}}]{enoch06}
{Enoch}, M.~L., {Young}, K.~E., {Glenn}, J., {et~al.} 2006, \apj, 638, 293

\bibitem[{{Evans} {et~al.}(2003){Evans}, {Allen}, {Blake}, {Boogert}, {Bourke},
  {Harvey}, {Kessler}, {Koerner}, {Lee}, {Mundy}, {Myers}, {Padgett},
  {Pontoppidan}, {Sargent}, {Stapelfeldt}, {van Dishoeck}, {Young}, \&
  {Young}}]{c2d}
{Evans}, N.~J., {Allen}, L.~E., {Blake}, G.~A., {et~al.} 2003, \pasp, 115, 965

\bibitem[{{Ferreira} \& {Pelletier}(1995)}]{ferreira95}
{Ferreira}, J. \& {Pelletier}, G. 1995, \aap, 295, 807

\bibitem[{{Frerking} {et~al.}(1982){Frerking}, {Langer}, \&
  {Wilson}}]{frerking82}
{Frerking}, M.~A., {Langer}, W.~D., \& {Wilson}, R.~W. 1982, \apj, 262, 590

\bibitem[{{Fuller} \& {Ladd}(2002)}]{fullerladd02}
{Fuller}, G.~A. \& {Ladd}, E.~F. 2002, \apj, 573, 699

\bibitem[{{Giannini} {et~al.}(2001){Giannini}, {Nisini}, \&
  {Lorenzetti}}]{giannini01}
{Giannini}, T., {Nisini}, B., \& {Lorenzetti}, D. 2001, \apj, 555, 40

\bibitem[{{Gooch}(1996)}]{gooch96}
{Gooch}, R. 1996, in ASP Conf. Ser. 101: Astronomical Data Analysis Software
  and Systems V, 80

\bibitem[{{Greene} {et~al.}(1994){Greene}, {Wilking}, {Andre}, {Young}, \&
  {Lada}}]{greene94}
{Greene}, T.~P., {Wilking}, B.~A., {Andre}, P., {Young}, E.~T., \& {Lada},
  C.~J. 1994, \apj, 434, 614

\bibitem[{{Hatchell} {et~al.}(2006){Hatchell}, {Fuller}, {Richer}, {Harries},
  \& {Ladd}}]{class}
{Hatchell}, J., {Fuller}, G., {Richer}, J., {Harries}, T., \& {Ladd}, E. 2006,
  in prep. (Paper II)

\bibitem[{{Hatchell} {et~al.}(1999{\natexlab{a}}){Hatchell}, {Fuller}, \&
  {Ladd}}]{hatchell99b}
{Hatchell}, J., {Fuller}, G.~A., \& {Ladd}, E.~F. 1999{\natexlab{a}}, \aap,
  346, 278

\bibitem[{{Hatchell} {et~al.}(1999{\natexlab{b}}){Hatchell}, {Fuller}, \&
  {Ladd}}]{hatchell99a}
{Hatchell}, J., {Fuller}, G.~A., \& {Ladd}, E.~F. 1999{\natexlab{b}}, \aap,
  344, 687

\bibitem[{{Hatchell} {et~al.}(2005){Hatchell}, {Richer}, {Fuller},
  {Qualtrough}, {Ladd}, \& {Chandler}}]{paperI}
{Hatchell}, J., {Richer}, J.~S., {Fuller}, G.~A., {et~al.} 2005, \aap, 440,
  151, (Paper~I)

\bibitem[{{Henriksen} {et~al.}(1997){Henriksen}, {Andre}, \&
  {Bontemps}}]{henriksen97}
{Henriksen}, R., {Andre}, P., \& {Bontemps}, S. 1997, \aap, 323, 549

\bibitem[{{Hogerheijde} {et~al.}(1998){Hogerheijde}, {van Dishoeck}, {Blake},
  \& {van Langevelde}}]{hogerheijde98}
{Hogerheijde}, M.~R., {van Dishoeck}, E.~F., {Blake}, G.~A., \& {van
  Langevelde}, H.~J. 1998, \apj, 502, 315

\bibitem[{{J{\o}rgensen} {et~al.}(2007){J{\o}rgensen}, {Johnstone}, {Kirk}, \&
  {Myers}}]{jorgensen07}
{J{\o}rgensen}, J.~K., {Johnstone}, D., {Kirk}, H., \& {Myers}, P.~C. 2007,
  \apj, 656, 293

\bibitem[{{Kenyon} \& {Hartmann}(1995)}]{kenyonhartmann95}
{Kenyon}, S.~J. \& {Hartmann}, L. 1995, \apjs, 101, 117

\bibitem[{{Kenyon} {et~al.}(1990){Kenyon}, {Hartmann}, {Strom}, \&
  {Strom}}]{kenyon90}
{Kenyon}, S.~J., {Hartmann}, L.~W., {Strom}, K.~M., \& {Strom}, S.~E. 1990,
  \aj, 99, 869

\bibitem[{{Kirk} {et~al.}(2006){Kirk}, {Johnstone}, \& {Di Francesco}}]{kirk06}
{Kirk}, H., {Johnstone}, D., \& {Di Francesco}, J. 2006, \apj, 646, 1009

\bibitem[{{Knee} \& {Sandell}(2000)}]{kneesandell00}
{Knee}, L.~B.~G. \& {Sandell}, G. 2000, \aap, 361, 671

\bibitem[{{Kregel} {et~al.}(2004){Kregel}, {van der Kruit}, \& {de
  Blok}}]{renzogram}
{Kregel}, M., {van der Kruit}, P.~C., \& {de Blok}, W.~J.~G. 2004, \mnras, 352,
  768

\bibitem[{{Lada}(1985)}]{lada85}
{Lada}, C.~J. 1985, \araa, 23, 267

\bibitem[{{Mangum} {et~al.}(1998){Mangum}, {Bontemps}, \& {Andre}}]{mangum98}
{Mangum}, J.~G., {Bontemps}, S., \& {Andre}, P. 1998, in 3rd Cologne-Zermatt
  Symposium, ed. V.~{Ossenkopf} ({Shaker-Verlag Aachen})

\bibitem[{{Margulis} \& {Lada}(1985)}]{margulis85}
{Margulis}, M. \& {Lada}, C.~J. 1985, \apj, 299, 925

\bibitem[{{Nisini} {et~al.}(2000){Nisini}, {Benedettini}, {Giannini},
  {Codella}, {Lorenzetti}, {di Giorgio}, \& {Richer}}]{nisini00}
{Nisini}, B., {Benedettini}, M., {Giannini}, T., {et~al.} 2000, \aap, 360, 297

\bibitem[{{Parker} {et~al.}(1991){Parker}, {Padman}, \&
  {Scott}}]{parkerpadman91}
{Parker}, N.~D., {Padman}, R., \& {Scott}, P.~F. 1991, \mnras, 252, 442

\bibitem[{{Pelletier} \& {Pudritz}(1992)}]{pelletier92}
{Pelletier}, G. \& {Pudritz}, R.~E. 1992, \apj, 394, 117

\bibitem[{{Raga} {et~al.}(1990){Raga}, {Binette}, {Canto}, \&
  {Calvet}}]{raga90}
{Raga}, A.~C., {Binette}, L., {Canto}, J., \& {Calvet}, N. 1990, \apj, 364, 601

\bibitem[{{Raga} {et~al.}(2004){Raga}, {Riera}, {Masciadri}, {Beck},
  {B{\"o}hm}, \& {Binette}}]{raga04}
{Raga}, A.~C., {Riera}, A., {Masciadri}, E., {et~al.} 2004, \aj, 127, 1081

\bibitem[{{Richer} {et~al.}(2000){Richer}, {Shepherd}, {Cabrit}, {Bachiller},
  \& {Churchwell}}]{richer00}
{Richer}, J.~S., {Shepherd}, D.~S., {Cabrit}, S., {Bachiller}, R., \&
  {Churchwell}, E. 2000, Protostars and Planets IV, 867

\bibitem[{{Rodriguez} {et~al.}(1997){Rodriguez}, {Anglada}, \&
  {Curiel}}]{rac97}
{Rodriguez}, L.~F., {Anglada}, G., \& {Curiel}, S. 1997, \apjl, 480, L125+

\bibitem[{{Shu} {et~al.}(1994){Shu}, {Najita}, {Ruden}, \& {Lizano}}]{shu94}
{Shu}, F.~H., {Najita}, J., {Ruden}, S.~P., \& {Lizano}, S. 1994, \apj, 429,
  797

\bibitem[{{Smith}(2000)}]{smith00}
{Smith}, M.~D. 2000, Irish Astronomical Journal, 27, 25

\bibitem[{{{\v C}ernis} \& {Strai{\v z}ys}(2003)}]{cernisstraizys03}
{{\v C}ernis}, K. \& {Strai{\v z}ys}, V. 2003, Baltic Astronomy, 12, 301

\bibitem[{{Walawender} {et~al.}(2006){Walawender}, {Bally}, {Kirk},
  {Johnstone}, {Reipurth}, \& {Aspin}}]{walawender06}
{Walawender}, J., {Bally}, J., {Kirk}, H., {et~al.} 2006, \aj, 132, 467

\bibitem[{{Wardle} \& {Koenigl}(1993)}]{wardle93}
{Wardle}, M. \& {Koenigl}, A. 1993, \apj, 410, 218

\bibitem[{{Wilking} {et~al.}(1989){Wilking}, {Lada}, \& {Young}}]{wilking89}
{Wilking}, B.~A., {Lada}, C.~J., \& {Young}, E.~T. 1989, \apj, 340, 823

\bibitem[{{Wilson} \& {Rood}(1994)}]{wilsonrood94}
{Wilson}, T.~L. \& {Rood}, R. 1994, \araa, 32, 191

\bibitem[{{Zhang} {et~al.}(2005){Zhang}, {Hunter}, {Brand}, {Sridharan},
  {Cesaroni}, {Molinari}, {Wang}, \& {Kramer}}]{zhang05}
{Zhang}, Q., {Hunter}, T.~R., {Brand}, J., {et~al.} 2005, \apj, 625, 864

\end{thebibliography}

\Online

\begin{figure*}[p]
\centering
\includegraphics[scale=0.9,angle=0]{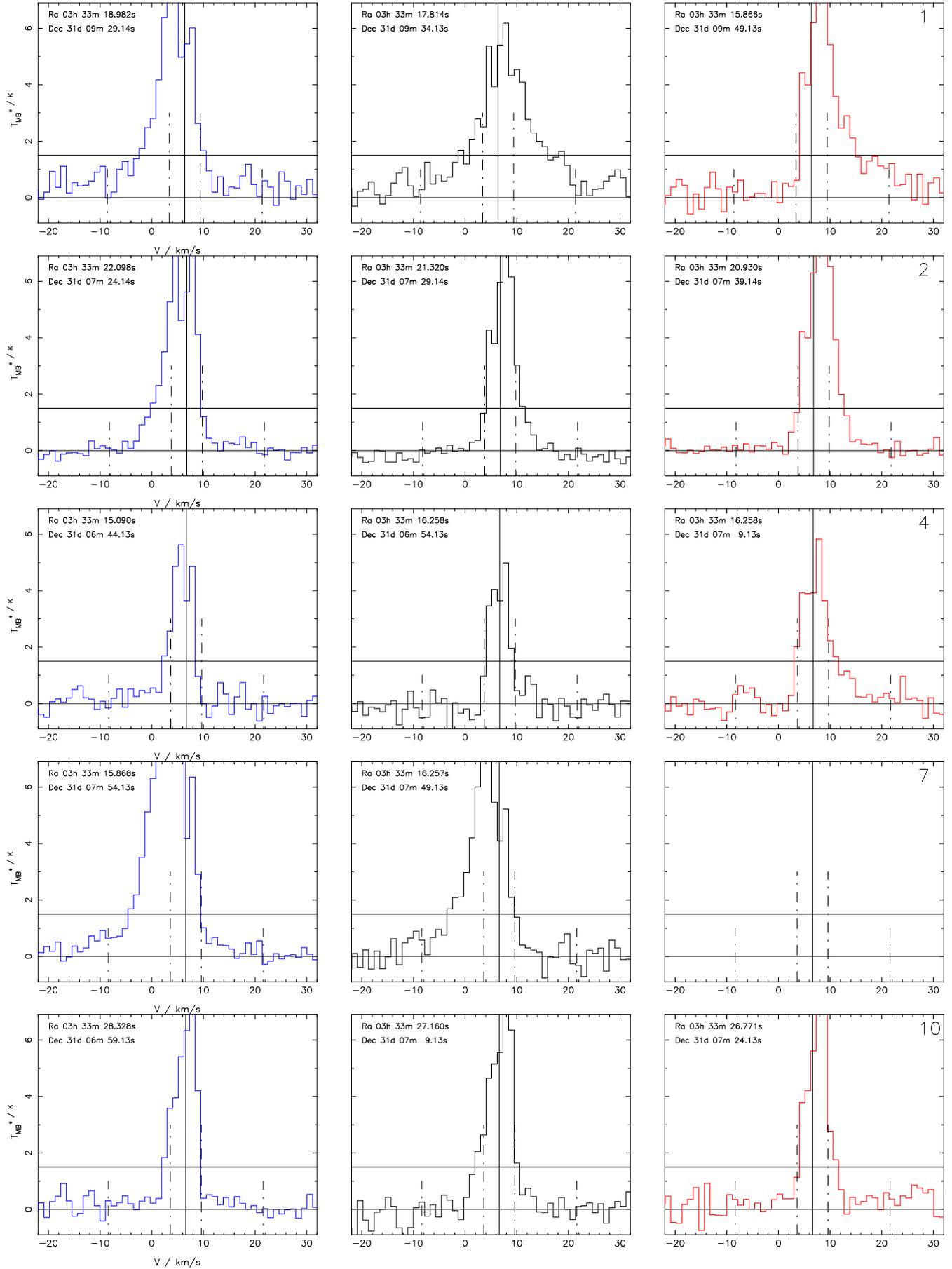}\\
\caption{ $^{12}$CO~3--2 spectra for outflow sources showing linewings at the 1.5~K ($3\sigma$) level at least 3~km~s$^{-1}$ from the ambient velocity (marked by inner solid and vertical lines).  The outer vertical lines mark the range of velocities included in the momentum flux calculation.}
\label{fig:outflowspec}
\end{figure*}
\newpage
\addtocounter{figure}{-1}
\begin{figure*}[p]
\centering
\includegraphics[scale=0.80,angle=0.0]{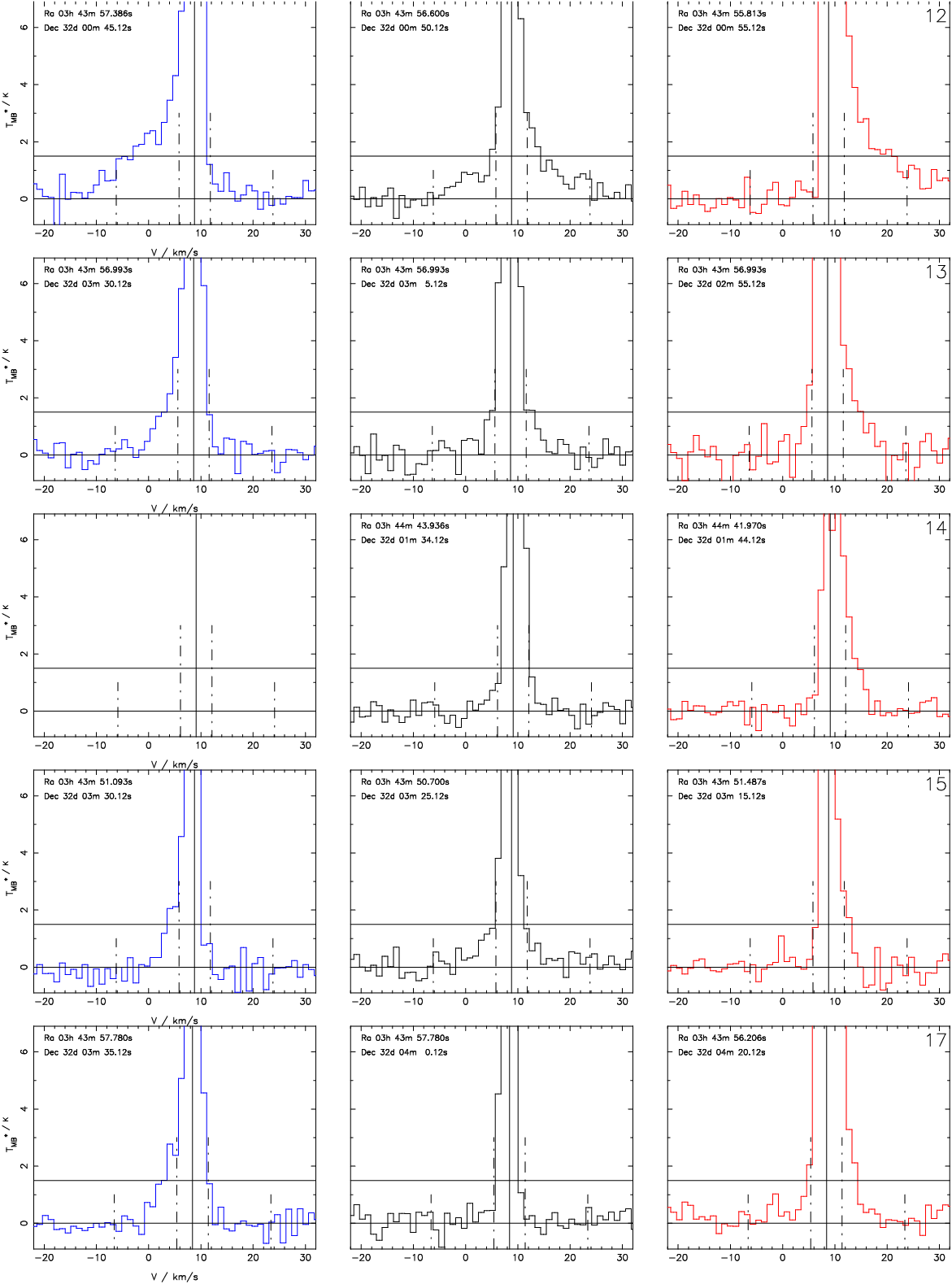}
\caption{{\bf continued.}}
\end{figure*}

\newpage
\addtocounter{figure}{-1}
\begin{figure*}[p]
\centering
\includegraphics[scale=0.80,angle=0.0]{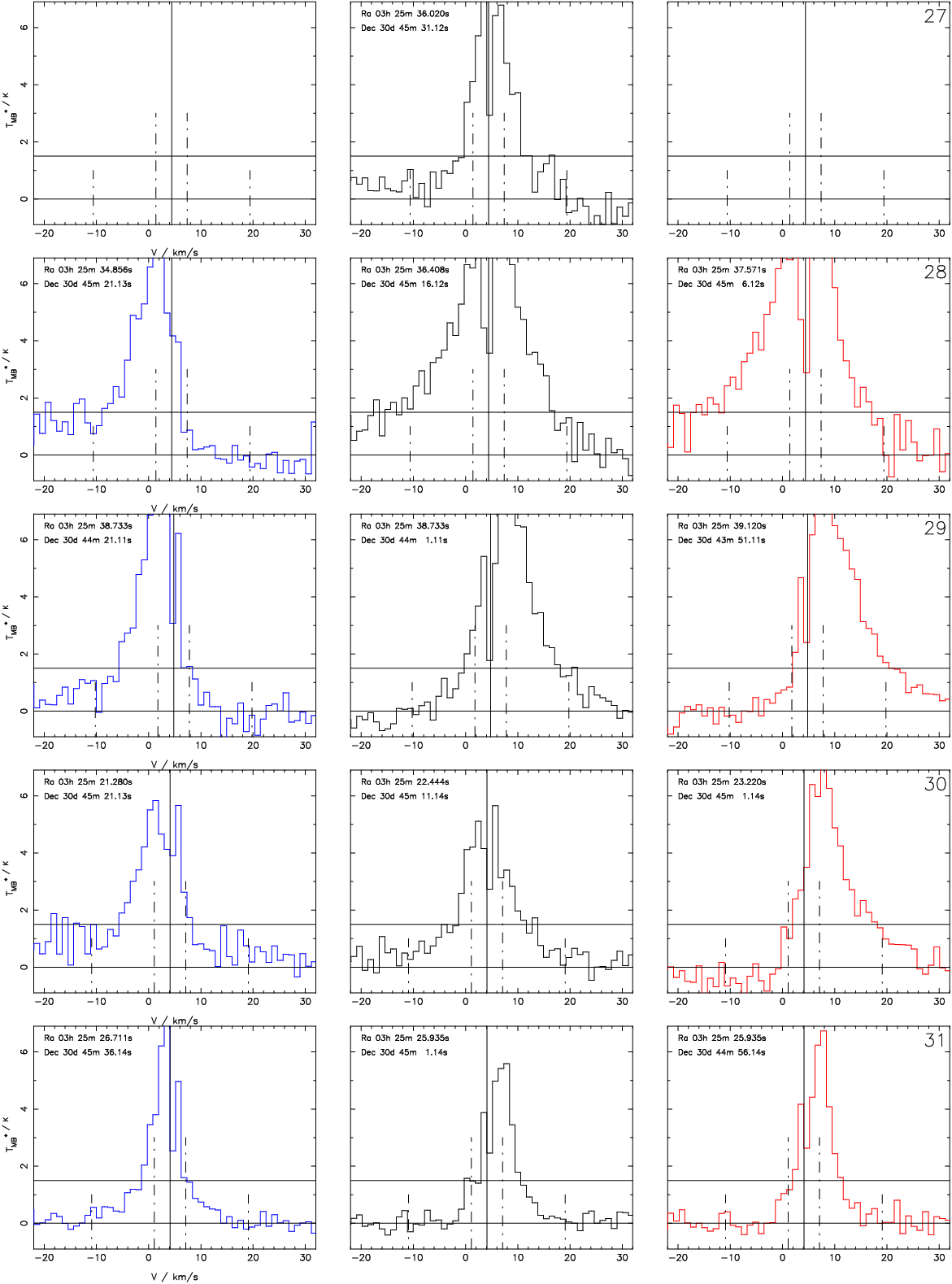}
\caption{{\bf continued.}}
\end{figure*}

\newpage
\addtocounter{figure}{-1}
\begin{figure*}[p]
\centering
\includegraphics[scale=0.80,angle=0.0]{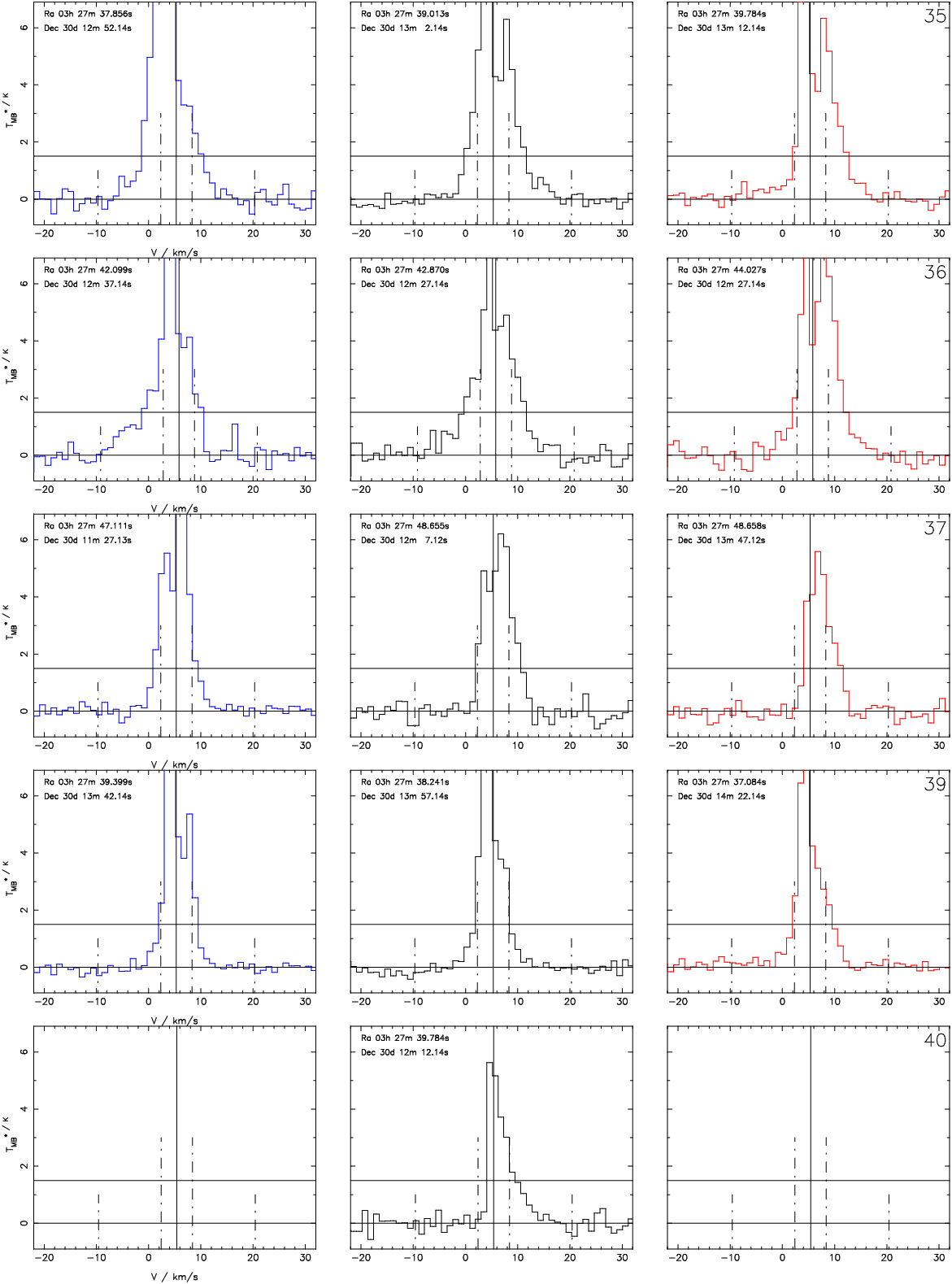}
\caption{{\bf continued.}}
\end{figure*}

\newpage
\addtocounter{figure}{-1}
\begin{figure*}[p]
\centering
\includegraphics[scale=0.80,angle=0.0]{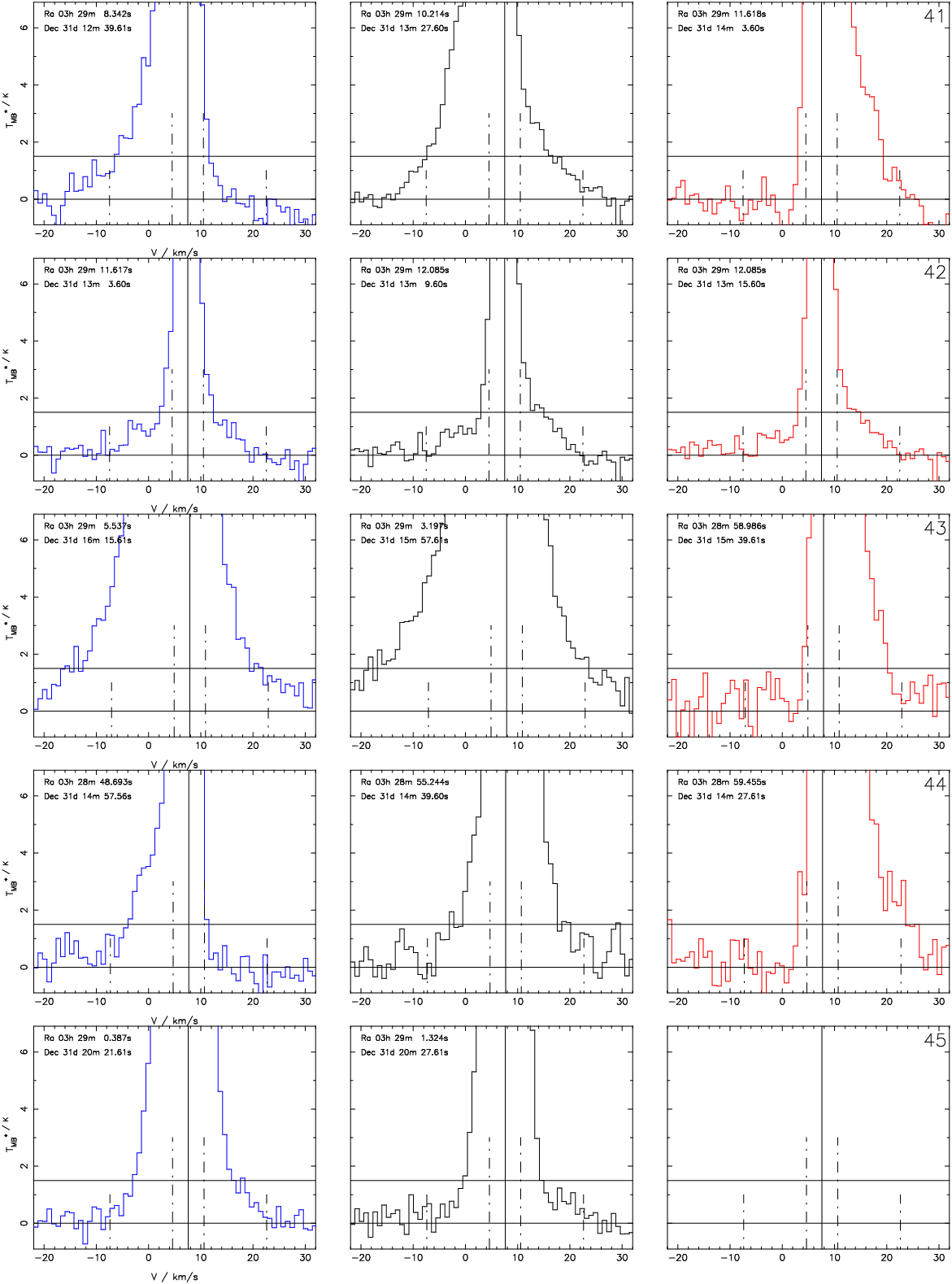}
\caption{{\bf continued.}}
\end{figure*}

\newpage
\addtocounter{figure}{-1}
\begin{figure*}[p]
\centering
\includegraphics[scale=0.80,angle=0.0]{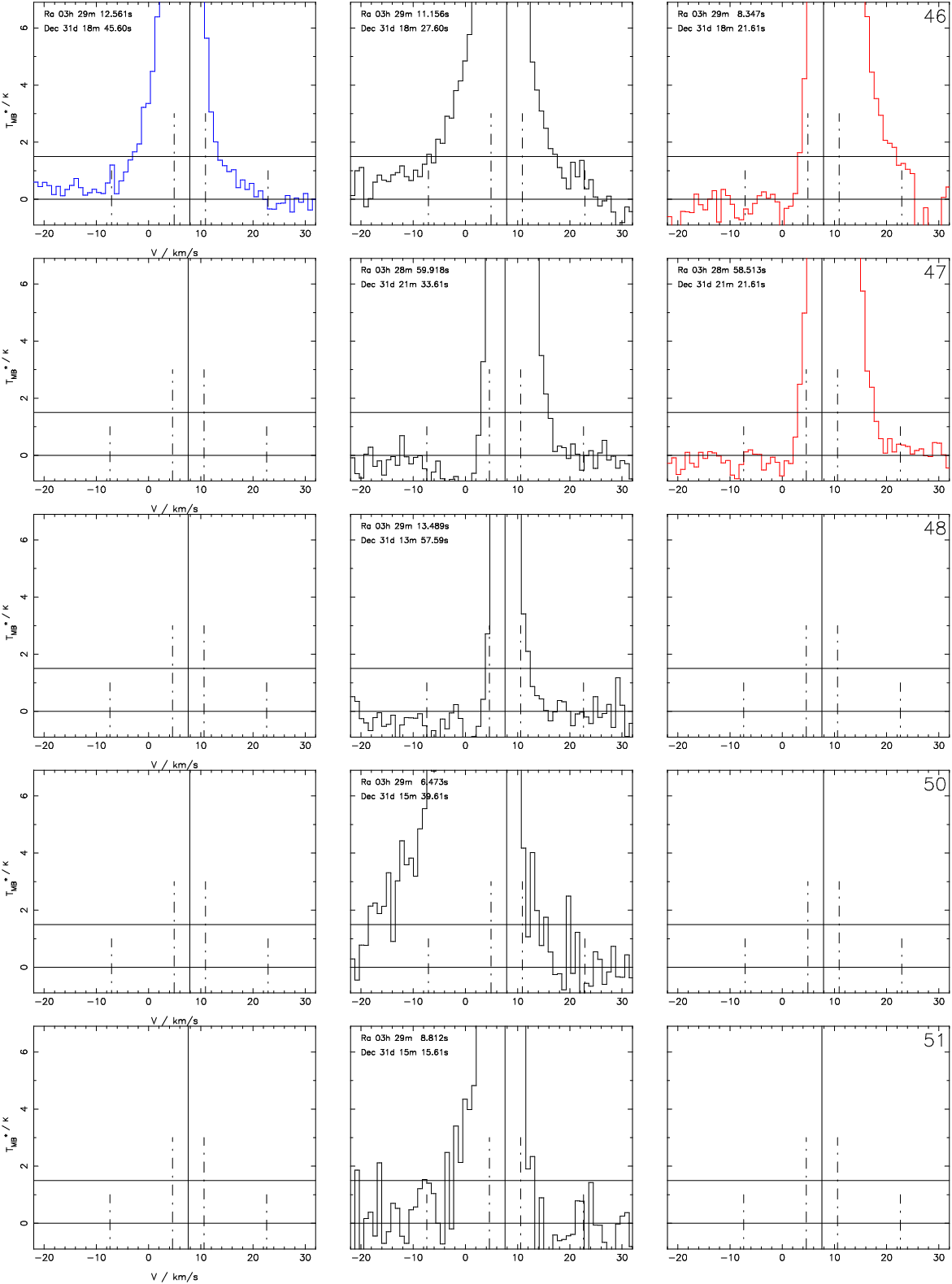}
\caption{{\bf continued.}}
\end{figure*}

\newpage
\addtocounter{figure}{-1}
\begin{figure*}[p]
\centering
\includegraphics[scale=0.80,angle=0.0]{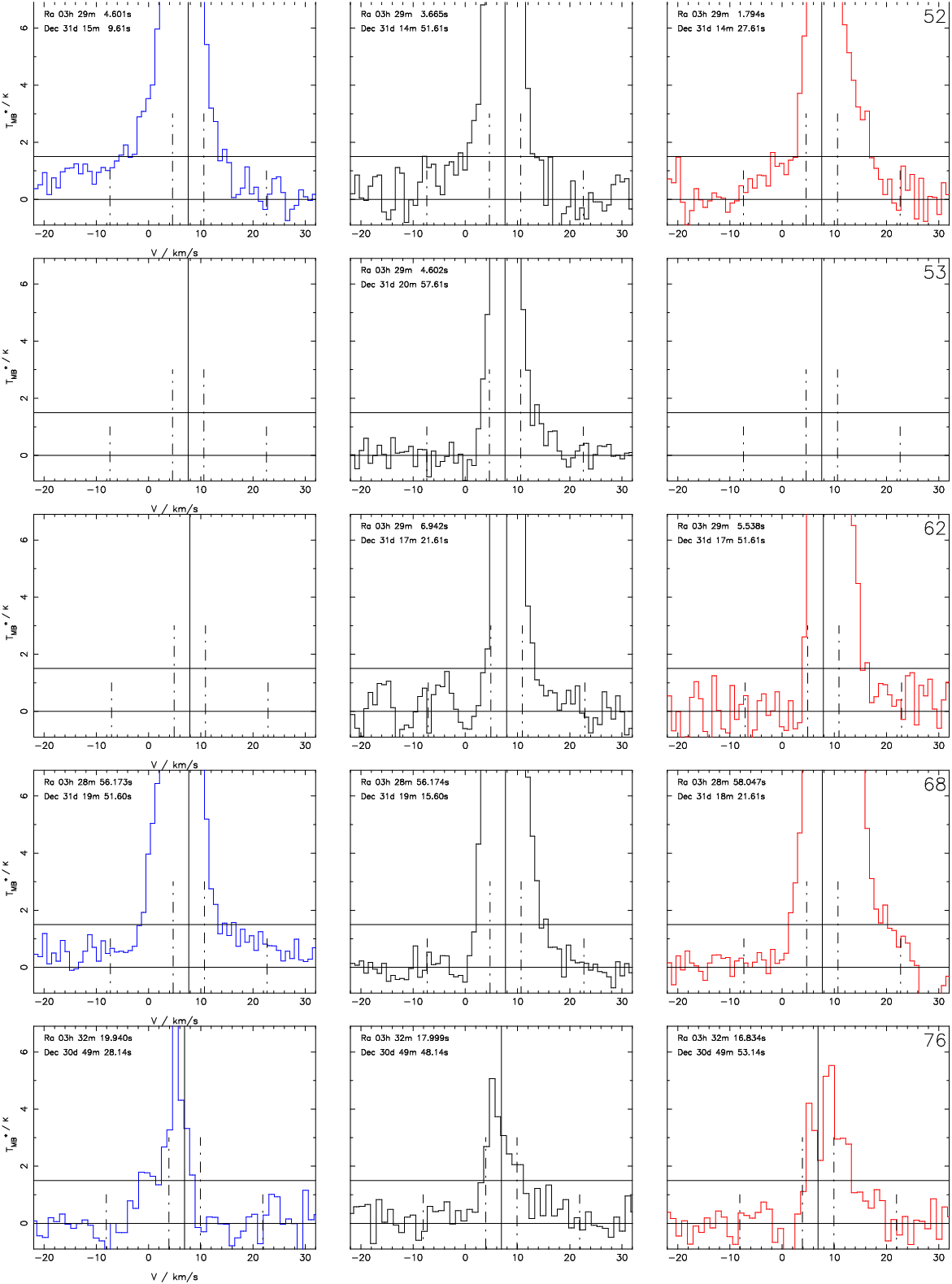}
\caption{{\bf continued.}}
\end{figure*}

\newpage
\addtocounter{figure}{-1}
\begin{figure*}[p]
\centering
\includegraphics[scale=0.80,angle=0.0]{outflowspec8.ps}
\caption{{\bf continued.}}
\end{figure*}

\begin{figure*}[p]
\centering
\includegraphics[scale=0.9,angle=0]{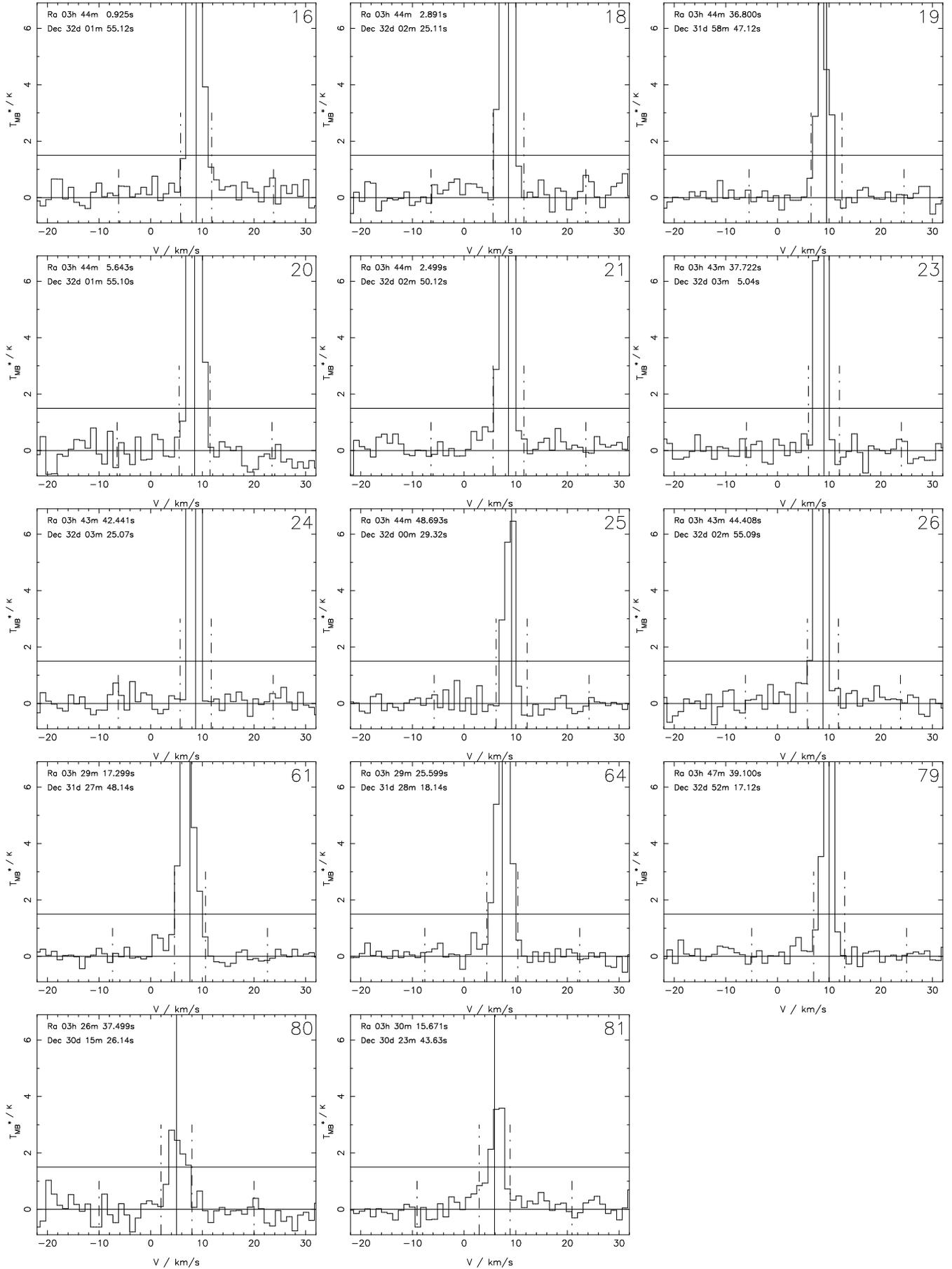}\\
\caption{ $^{12}$CO~3--2 spectra for sources with no outflow, showing no linewings above the 1.5~K ($3\sigma$) level (upper horizontal solid line) at  3~km~s$^{-1}$ from the ambient velocity (inner solid and vertical lines).}
\label{fig:nooutflowspec}
\end{figure*}

\appendix

\section{Calculation of outflow fraction probability}
\label{app:outflowprob}

Two measured quantities are involved in calculating the fraction of
protostars (outflow sources) in each mass bin.  The first is the
observed fraction of outflow sources in each mass bin m, $B = b/n$ (n
is the total observed sources in the bin).  The second is the
detection probability in that mass bin calculated from the simulation:
$A = a/N$ (N is the total simulated sources in the bin).  Both $a$ and
$b$ are expected to follow a binomial distribution so we can easily
calculate the corresponding uncertainties on $B$ and $C$: they are
just $\Bin(a,n,A)$ and $\Bin(b,n,B)$ respectively, where $\Bin(k,n,p)$
is the binomial distribution for $k$ successes from $n$ trials with
probability $p$.  With our estimates for the detected outflow fraction
$A$ and detection probability $B$ we can then estimate the true
protostellar fraction in the bin $C = B/A$.

Given that $a$ and $b$ are binomially distributed, we can also calculate the
probability distribution for $C=A/B$ by integrating the probability space $P(A,B)$ where $B = A/C$ is satisfied.  

\begin{eqnarray}
P(C) &=& \int\int P(A,B)\, \delta(C-A/B)\, dA dB \nonumber\\
     &=& \int P(A)\, P(B=A/C) dA\nonumber\\
     &=& \int \Bin(a,n,A)/n \times \Bin(b,n,B)/N \,dA\nonumber
\end{eqnarray}

This we integrated numerically for a range of values of $C$ for each
mass bin and the resulting probability distributions are plotted as
greyscale in Fig.~\ref{fig:outflowstats}.

\end{document}